\documentclass[11pt,preprint,flushrt]{aastex}
\usepackage{amsmath}
\allowdisplaybreaks[1]

\def\eqq#1{Equation~(\ref{#1})}
\newcommand\etal{{\it et al.\/}}
\newcommand\cf{{\it cf.\ }}
\newcommand\eg{{\it e.g.\ }}

\newcommand\etc{{\it etc.\/}}

\newcommand{\bfS}{\mbox{\boldmath $\Sigma$}}
\newcommand{\bfe}{\mbox{\bf e}}
\newcommand{\bfd}{\mbox{\boldmath $\delta$}}
\newcommand{\bft}{\mbox{\boldmath $\theta$}}
\newcommand{\map}{\mbox{$M_{\rm ap}$}}
\newcommand{\mx}{\mbox{$M_{\times}$}}
\newcommand{\mxsq}{\mbox{$\langle M_{\times}^2\rangle$}}
\newcommand{\mapsq}{\mbox{$\langle M_{\rm ap}^2\rangle$}}
\newcommand{\gamsq}{\mbox{$\langle \gamma^2\rangle$}}

\begin{document}

\title{Weak Lensing Results from the 75 Square Degree CTIO Survey}

\author{M. Jarvis\altaffilmark{1,2}, G. M. Bernstein\altaffilmark{1,2}, 
P. Fischer\altaffilmark{1,3}, D. Smith\altaffilmark{1,4}}
\affil{Department of Astronomy, University of Michigan,
830 Dennison Bldg., Ann Arbor, MI 48109}
\author{B. Jain}
\affil{Dept. of Physics and Astronomy, University of Pennsylvania,
Philadelphia, PA 19104}
\author{J. A.  Tyson\altaffilmark{1}, and D. Wittman\altaffilmark{1}}
\affil{Bell Laboratories, Lucent Technologies, Murray Hill, NJ 07974}
\email{mjarvis, garyb, bjain@physics.upenn.edu,
tyson, wittman@physics.bell-labs.com}

\altaffiltext{1}{Visiting Astronomer, National Optical Astronomy
Observatories, which is operated by the Association of Universities
for Research in Astronomy, Inc., under contract to the National
Science Foundation.}
\altaffiltext{2}{Current address: Dept. of Physics and Astronomy,
University of Pennsylvania, Philadelphia, PA 19104}
\altaffiltext{3}{Current Address: Bank of Nova Scotia, Trade Floor
Risk Management, 40 King St. W., Toronto ON}
\altaffiltext{4}{Current Address: Glenelg High School, Glenelg MD 21737}

\begin{abstract}
We measure seeing-corrected ellipticities for $2\times10^6$ galaxies
with magnitude $R\le23$ in 12 widely separated fields totalling
75~deg$^2$ of sky.  At angular scales $\gtrsim30\arcmin$, ellipticity
correlations are detected at high significance and exhibit nearly the pure
``E-mode'' behavior expected of weak gravitational lensing.  
Even when smoothed to the full field size of 2\fdg5, 
which is $\approx25h^{-1}$~Mpc at the lens distances,
an rms shear variance
of $\langle \gamma^2 \rangle^{1/2}=0.0012\pm0.0003$ is detected.
At smaller angular scales there is significant ``B-mode'' power, an
indication of residual uncorrected PSF distortions. 
The $>30\arcmin$ data constrain the power spectrum of
matter fluctuations on comoving scales of $\approx10h^{-1}$~Mpc  to
have 
$\sigma_8 (\Omega_m/0.3)^{0.57} = 0.71^{+0.12}_{-0.16}$
(95\% CL, $\Lambda$CDM, $\Gamma=0.21$),
where the systematic error includes statistical and calibration
uncertainties, cosmic variance, and a conservative estimate of
systematic contamination based upon the detected B-mode signal.
This normalization of the power spectrum is lower than, but 
generally consistent with, previous weak-lensing results, is
at the lower end of the $\sigma_8$ range from various analyses of
galaxy cluster abundances, and agrees with recent determinations from
CMB and galaxy clustering.

The large and dispersed
sky coverage of our survey reduces random errors and cosmic variance, while the
relatively shallow depth allows us to use existing redshift-survey
data to reduce systematic uncertainties in the $N(z)$ distribution to
insignificance.   
Reanalysis of the data with more sophisticated
algorithms will hopefully reduce the systematic (B-mode)
contamination, and allow more precise, multidimensional  constraint of
cosmological parameters. 

\end{abstract}

\keywords{gravitational lensing; cosmology; large-scale structure}

\section{Introduction}

The realization that weak gravitational lensing effects could reveal
the power spectrum of matter fluctuations in the Universe
\citep{Va83,Mi91,Ka92} strongly motivates large-scale imaging
surveys of faint galaxies.  Weak gravitational lensing constrains
fundamental parameters under the current paradigm that the power
spectrum of matter evolves from primordial fluctuations due to
gravitational instability.  Comparison of weak lensing power at
$z\sim0$ with measurements of the cosmic background anisotropy at
$z=1000$ can ultimately test this underlying paradigm to high
precision.

The coherent distortions induced by weak lensing are, however, lost in
the intrinsic shape variations of the source galaxies unless a very
large number of galaxy shapes can be determined to beat down this
``shape noise.''  This was the primary impetus behind the construction
of the Big Throughput Camera (BTC; \citet{Wi98}) and other
high-efficiency CCD mosaic cameras.  We report here the results of a
large weak-lensing survey conducted with the BTC camera and its
successor, the NOAO Mosaic II imager \citep{Mu98}.

Firm detections of weak lensing in random fields were first reported
using early data from mosaic ground-based cameras 
\citep{Wi00,KWL00,vW00,Bac00}, HST \citep{Rh00},
and single-CCD cameras \citep{Bac00, Ma01}.
These initial efforts did not place strong constraints on the matter
spectrum.  This is due in part to their relatively small number of galaxy
samples ($\approx 10^5$ or fewer).  Close inspection of these data
also reveal, though, that the methods that were used to remove systematic
distortions induced by PSF ellipticities have left residual signals
that may contaminate the lensing observations.

More recently, \citet{Ho02}[HYG02], \citet{vW02}[vW02],
\citet{Bac02}, and \citet{Re02}, have derived
more precise 
constraints on $\Omega_m$ and the normalization $\sigma_8$ of the
matter power spectrum by analyzing larger samples of galaxy shapes.
In this paper we present 
constraints on the power spectrum from the largest weak-lensing survey to 
date, using 75~deg$^2$ of images collected with the BTC and Mosaic II imagers.
HYG02, vW02, and this paper not only have large sky coverage, but
also make use of the techniques presented by \citet{Cr02,
Sch02}; and \citet{Pen02} for distinguishing ``E-mode'' distortion
patterns, which should be produced by lensing, from ``B-mode'' power,
which indicates the presence of uncorrected systematic errors.  The E/B
decomposition provides an important validity check as well as
improving the $S/N$ ratio by $\sqrt 2$.  Our program is
distinguished by relying exclusively on
galaxies with magnitude $R<23$, for which the redshift distribution
$N(z)$ is well measured by spectroscopic redshift surveys.  The
accurate calibration eliminates another source of systematic error.
Our results differ as well by making use of
many of the techniques developed in \citet{BJ02}[BJ02] for extraction of
galaxy ellipticities and lensing distortion signals in the face of
asymmetric PSFs.  Most other results to date make use the formalism of
\citet{KSB} and its modifications.  Our data, reductions
methods, and survey depth are largely disjoint from other authors'.
Given the subtlety of the weak
lensing measurements, we seek reassurance that independent methods
yield similar cosmological results.

\section{Data}
\label{data}

\subsection{Observations}
\label{observations}

The data were taken using the Blanco 4 meter telescope at
Cerro Tololo Interamerican Observatory (CTIO) in Chile from December,
1996 to July, 2000.  The telescope changed its wide-field imager 
from the BTC to the Mosaic II in 1999, and approximately half of the
data were taken using each camera.  This is beneficial to 
us, since some systematic effects are presumably different for the 
two cameras, so we can compare the results from each 
camera.  Table~\ref{obstable} summarizes the observing runs.

\begin{deluxetable}{ccc}
\tablewidth{0pt}
\tablecaption{Summary of Observations}
\tablehead{
\colhead{Date} &
\colhead{Clear Nights} &
\colhead{Camera}
}
\startdata
1996 Dec & 3\tablenotemark{1} & BTC \\
1997 Feb & 3\tablenotemark{1} & BTC \\
1998 Nov & 4 & BTC \\
1999 Feb & 5 & BTC \\
2000 Jan & 6 & Mosaic \\
2000 Jul & 4 & Mosaic 
\enddata
\tablenotetext{1}{Less than half of the time in these runs was devoted to
this project.}
\label{obstable}
\end{deluxetable}

We observed 12 fields, each approximately 2\fdg5 square.  
Table~\ref{fieldstable} summarizes the locations, area, and camera for each 
field.  We also list the galactic extinction, $A_R$ 
(\cf step~\ref{stepphotometry}), the mean distortion for each field 
(\cf \S\ref{overallshear}) and the total number of galaxies used for
shape measurement after the cuts described in \S\ref{datareduction}.
Note that we were unable to finish two of the fields (K and R).  
These two are approximately half the full height in declination, but still
give us useful statistics on scales of 2\fdg5 in the RA direction.

Each sky location is observed in three distinct 5-minute exposures,
all in the $R$ band.
Exposure pointings are taken in an interlaced pattern that places each
galaxy's image on two or three different CCDs of the mosaic.  This
makes it easier to detect systematic errors that depend upon chip
location, as discussed further in \S\ref{galshapevschip}.
The large dithers also allow us to eliminate chip defects, scattered
light features, and ghost images from our object catalogs by
requiring galaxies to have multiple coincident detections. 
Each BTC field has 112 exposures with 4 $2048^2$ CCDs per
exposure, with a scale of 0\farcs43 per pixel.  Each Mosaic field has 75
exposures, 8 $2048\times4096$ CCDs per exposure,\footnote{
During the July, 2000 run, two of the chips failed.  One was
out for two nights and the other for three nights.  so
more exposures were taken in some fields to compensate for the lost 
area per exposure.}
and 0\farcs27 per
pixel.  In all, the dataset contains 1155 exposures, with
$>6400$ CCD images.  We clearly want to avoid the need to examine the images
or catalogs by eye at any point in the reduction process.

The magnitude at which galaxy completeness drops to 50\% is 
$R\approx23.5$, and varies somewhat from field to field due to differences
in seeing and other factors (\cf Figure~\ref{weightvsmagplot}).  
We use galaxies with
$19<R<23$ in order to have an approximately consistent depth for all fields.
The median seeing for all the exposures is 1.05 arcsec FWHM, 
with most of the exposures between 0.9 and 1.3 arcsec.  

\begin{deluxetable}{crrcccccc}
\tablewidth{0pt}
\tablecaption{Summary of Fields Observed}
\tablehead{
\colhead{Label} &
\colhead{RA} &
\colhead{Dec} &
\colhead{Area} &
\colhead{Camera} &
\colhead{$A_R$} &
\multicolumn{2}{c}{Mean Distortion (\%)} &
\colhead{$N_{gal}$} \\
\colhead{} &
\multicolumn{2}{c}{(J2000)} &
\colhead{(deg$^2$)} &
\colhead{} &
\colhead{(mag)} &
\colhead{$\delta_+$} & 
\colhead{$\delta_\times$} & 
\colhead{($\times 10^3$)}
}
\startdata
K & $02^{\rm h} 27^{\rm m}$ & $-00\arcdeg25\arcmin$ & 4.2 & Mixed & 0.080 & $-0.03$ & $-0.37$ &123\\
H & $03^{\rm h} 55^{\rm m}$ & $-42\arcdeg00\arcmin$ & 6.8 & BTC & 0.018 & $+0.33$ & $+0.11$ &160\\
J & $03^{\rm h} 58^{\rm m}$ & $-32\arcdeg53\arcmin$ & 6.8 & Mixed & 0.022 & $+0.21$ & $+0.03$ &184\\
N & $05^{\rm h} 21^{\rm m}$ & $-30\arcdeg11\arcmin$ & 6.8 & BTC & 0.045 & $-0.33$ & $+0.37$ &154\\
A & $10^{\rm h} 07^{\rm m}$ & $-05\arcdeg48\arcmin$ & 6.8 & BTC & 0.118 & $-0.01$ & $+0.13$ &168\\
M & $10^{\rm h} 26^{\rm m}$ & $-11\arcdeg34\arcmin$ & 6.8 & BTC & 0.121 & $-0.07$ & $-0.35$ &163\\
Q & $10^{\rm h} 41^{\rm m}$ & $-20\arcdeg48\arcmin$ & 6.8 & Mosaic & 0.106 & $+0.37$ & $+0.01$ &188\\
L & $12^{\rm h} 01^{\rm m}$ & $-11\arcdeg51\arcmin$ & 6.8 & Mixed & 0.133 & $-0.21$ & $+0.21$ &179\\
T & $14^{\rm h} 25^{\rm m}$ & $-00\arcdeg19\arcmin$ & 6.8 & Mosaic & 0.111 & $-0.21$ & $-0.10$ &142\\
X & $21^{\rm h} 39^{\rm m}$ & $-41\arcdeg07\arcmin$ & 6.8 & Mosaic & 0.056 & $+0.32$ & $-0.31$ &148\\
R & $21^{\rm h} 52^{\rm m}$ & $-31\arcdeg17\arcmin$ & 3.4 & Mosaic & 0.062 & $-0.21$ & $-0.42$ &\phn91\\
G & $23^{\rm h} 54^{\rm m}$ & $-42\arcdeg11\arcmin$ & 6.8 & Mixed & 0.031 & $+0.00$ & $-0.16$ &150
\enddata
\label{fieldstable}
\end{deluxetable}

\subsection{Image Reduction}
\label{datareduction}

Extracting useful shapes from observations of distant galaxies is a
complicated procedure. The CTIO Survey data have been analyzed using a
subset of the techniques described in BJ02. We list below the steps
involved in our reduction process, and refer to BJ02 for
a more complete description of each step.  At no point in the
reduction do we sum exposures:  all measurements are made on
individual exposures, with merging of exposures occurring at the
catalog level.  This is ``Method 1'' described in BJ02 \S4.

\begin{enumerate}
\item {\em Bias Subtraction and Flat Fielding}
\label{stepbiasflat}

We do this in the normal way using the IRAF packages CCDRED
and MOSRED\footnote{
IRAF is distributed by NOAO, which is operated by AURA under 
cooperative agreement with NSF.}.
Note that 
our observing scheme allows us to make excellent dark sky flats, since we 
move substantially after every exposure.

\item {\em Object Detection}
\label{stepsex}

We use the program {\tt SExtractor} \citep{SEx} for the initial object 
detection.  As discussed in BJ \S8.1, any surface-brightness threshold
detection scheme will result in a 
selection bias, whereby objects similar in shape to the PSF will be more
likely to be detected.  We set the SExtractor detection threshold
very low, such that a significant fraction of the detections 
are noise, or unusably faint galaxies.  Later (step~\ref{stepuber}), 
we select from our list of objects according to a significance 
parameter that is unbiased by the PSF shape.  Requiring coincident
detections on more than one exposure eliminates the vast majority of
the noise detections and spurious objects.  Galaxy total magnitudes
and sky levels are obtained from the SExtractor catalogs.

\item {\em Field Registration and Distortion Measurement}
\label{stepdistortion}

Registration of the images 
is done in two stages.  In the first stage, we register each field by
matching the  
bright objects ($m < 19$) to the positions of stars given by the
USNO A2.0 catalog \citep{USNOA2}.  There are typically 100 or fewer
matching stars per CCD image, 
which is not sufficient to adequately fit
for all terms of the coordinate map.  

From the first stage coordinate maps, we can match all detections
of a given galaxy.  Each galaxy is typically observed 3 or 4 times,
and we ignore any object which is detected only once, most of which are
either noise or edge objects.  

In the second stage of the coordinate map
determination, we fit the full polynomial telescope distortion model
by minimizing the exposure-to-exposure variance in the positions of
multiply-detected galaxies.  The USNO star
positions are included in the fit to tie the solutions to the
astrometric frame.
Note that this stage of this procedure would be impossible 
if we had not dithered
our exposures by large amounts, because a solution with only small
shifts in galaxy positions leaves many of the distortion parameters
degenerate, and there are too few USNO stars to constrain all the
higher-order distortion terms.

\item {\em Photometry}
\label{stepphotometry}

Precise photometry is not extremely important for this study, since 
we are primarily concerned with the shapes of galaxies, not their
overall intensity.  We do, however, need reasonable
magnitude estimates for our galaxies to infer their distribution
of redshifts (see \S\ref{zdist}). For our $R<23$ sample, a magnitude
calibration error of $\Delta m$ leads to an erroneous factor
$10^{-0.04\Delta m}$ in the inferred $\sigma_8$.

We use the SExtractor ``best'' magnitudes as our photometry measurement.
This magnitude is then corrected for the Jacobian of the distortion 
map found for each image in step~\ref{stepdistortion}.
All exposures of a given field are put on a common photometric system
by minimizing the exposure-to-exposure variance of multiply-detected
galaxies' magnitudes.
This procedure only determines the relative magnitudes from image to image.
We obtain the overall zeropoint by measuring several \citet{Lan92}
stars for each run.  Our magnitudes are thus
tied to his Cousins $R$ filter system.  No significant color terms
are detected.
The uncertainty in the magnitudes from this zeropoint determination
is $\approx0.02$~mag, which is quite 
sufficient for our redshift calibration.

Finally, we correct for the galactic extinction in each field using the 
dust maps of \citet{Sc98} assuming a value of $A_R/E_{B-V} = 2.63$ for
our CTIO $R$ filter.  The extinction values used for each field are listed
in Table~\ref{fieldstable}, and the largest such correction (for field L) 
is 0.133 magnitudes.  

\item {\em Initial Shape Measurement}
\label{stepellipto}

The shape of a galaxy is described by two components of its ellipticity.
These $e_+$ and $e_\times$ quantities are determined for each galaxy
on each exposure using the algorithms in \S3 of BJ02, in which
ellipticities are defined from weighted central second moments of each
galaxy.  The weights are elliptical Gaussians, with size, shape, and
centroid iterated to match those of the galaxy.
If the iteration does not converge,
usually due to crowding by nearby objects or edge effects, then we
discard the galaxy.
These measurements are made in sky coordinates rather than in pixel 
coordinates to remove the effects of telescope distortion.

\item {\em Star Identification}
\label{stepfindstars}

We need to know the point-spread function (PSF) for each exposure, 
so we can remove its effects.
Thus we need to identify the stars in each exposure.  
The usual method is to use a size-magnitude scatter plot, 
and look for an arm in the distribution at small sizes which separates
from the main swath of galaxies.  There is often an arm at smaller size, 
corresponding to cosmic rays, which does not extend as far to bright
magnitudes as the stellar arm.

This identification is very easy to do by eye,
but it is not trivial to construct an algorithm to automate the process.
A full description of the algorithm we use is 
in a forthcoming paper \citep{Jar02}, however it might be of interest
to mention one of the particular difficulties of this analysis for
large field cameras.
The images from the BTC camera have significant astigmatism 
which enlarge the stars near the corners of the array.
(The seeing on the Mosaic images is more uniform, but still has 
some of this effect.)
Thus, the stars near the  
corners of the array do not have the same size as most of the
stars on the size-magnitude plot.  If this is not carefully
taken into account, the stars near the corners will be missed, and 
one will not be able to accurately describe the PSF shape across
the whole chip.  This will, of course, lead to significant systematic
errors in the final shapes, so an accurate and complete identification of 
the stars is very important.

\item {\em Image Convolution}
\label{stepconvolve}

Correction for the effects of the PSF on galaxy shapes is done in
two steps.  First we remove the ellipticity induced by the 
PSF by convolving with a kernel which makes all the stars look round, 
using the methods in \S7 of BJ02.  Then we correct 
for the dilution of the shapes analytically
according to the size of the PSF at the location of each galaxy
(in step~\ref{stepuber}), as described in Appendix~C of BJ02.

For each star identified in step~\ref{stepfindstars}, 
we find a $7\times7$ pixel kernel which makes the star round.  
In the language of the Laguerre decomposition described in BJ02 \S6.3, 
we find a kernel for which $b_{20}$, $b_{31}$, and $b_{42}$ are all 0.  
The most important term for insuring that the PSF does not bias the
galaxy shapes is $b_{20} \propto e_+ + ie_\times$. The
$b_{31}$ and $b_{42}$ are higher order terms which can also cause a bias 
in galaxy shapes, so we make them vanish as well.  
The methods for finding the appropriate kernel are described in BJ02
\S7. 

Once we have a kernel for each star in the image, we need to
interpolate this kernel across the image.  The kernel is described by
a set of coefficients, so we actually fit each coefficient across the image.  
This apparently simple interpolation task has been found by us and
other practitioners \citep{vW02,Bac02} to require great care to avoid
inducing spurious distortion power. With too little freedom, 
an interpolant misses real variation in some regions of the chip, 
but with too much freedom, the fit adds spurious wiggles in
other regions or at the chip edges.
We find that simple polynomial fits could not simultaneously avoid
both problems, especially on the BTC PSF structure.  We instead
use a smoothing spline algorithm developed by 
\citet{Gu91}, which is available from NetLib as Rkpack\footnote{
http://www.netlib.org/gcv/rkpk.shar}.

As noted in BJ02 \S7.5, the convolution is efficient since each of 
the $7\times7$ kernel components is the result of three successive
$3\times3$ kernel convolutions.

We want to avoid extrapolating the kernel into
regions where the PSF is not well constrained.
We hence reject all galaxies from our catalogs 
which are too far from any
valid PSF-template star or which are outside the bounding convex
polygon
of all such stars.  We also reject regions which are near 
extremely bright stars to avoid spurious detections due to diffraction
spikes, rings or bleeding columns.

\item {\em Remeasurement of Shapes}
\label{stepellipto2}

The convolved images now have stars that are round.  Therefore, the 
intrinsic sky image
has now effectively been convolved with a round PSF, 
rather than the elliptical PSF of the original images.
When we measure the shapes of galaxies on this image, we should have
shapes which are unbiased by the PSF.  The measurement again uses the
method of BJ02 \S3.

Note, however, that the shapes are not yet the true shapes of the
galaxies. A round PSF makes a galaxy 
appear more round than it really is.  This is called PSF dilution, since 
the magnitude of the ellipticity is reduced by the PSF.  
Appendix~C of BJ02 derives an approximate factor $R$ by which the
shape has been 
diluted.  Each galaxy shape must be corrected for dilution by the factor
$1/R$ to obtain an estimate of the true shape of the galaxy.  This is
done in step~\ref{stepuber}.

\item {\em Removal of Centroid Bias}
\label{removecentroidbias}

BJ \S8.2 describes the ``centroid bias.''  Essentially,
the centroids of galaxies are more uncertain in the direction of the PSF 
elongation than they are perpendicular to this direction. 
This anisotropic error in the centroid position affects the measured
shapes as well and leaves a bias in the shapes relative to the 
observed PSF shape.

The functional form of the bias is expected to be
\begin{equation}
e_{bias} = K \frac{\sigma^2_{\rm PSF} / \sigma^2_{\rm gal}}{\nu^2} e_{\rm PSF} 
\end{equation}
where $\nu$ is the significance of the galaxy detection 
(see step~\ref{stepuber}),
and the $\sigma^2$ are weighted second radial moments (sizes) of the PSF
and the galaxy.
We measure $K$ empirically simply by fitting all of the observed shapes to 
this form, obtaining the value $K\approx -20$ for
both BTC and Mosaic images.
The bias is then subtracted from the shape of each galaxy.

\item {\em Combination of Measurements}
\label{stepuber}

Step~\ref{stepdistortion} yields a list of multiply-detected galaxies
and the exposures on which each galaxy was measured.
We can now combine the set of shape measurements for each galaxy into
a single maximum likelihood estimate of the shape according to the
methods of BJ \S4.2. 

Before averaging, we correct for the dilution mentioned above
by dividing each shape by the dilution $R$ (\cf step~\ref{stepellipto2}).
Galaxy measurements with $R<0.1$ are rejected.

This step is also where we select our galaxies using an unbiased
estimate of the significance $\nu$.  As discussed by \citet{Ka00} and
in BJ02 \S8.1, a surface-brightness selection of galaxies (as for
SExtractor) is 
inherently  biased toward galaxies aligned with the PSF.  
Galaxies which are the same shape 
as the PSF are essentially matched-filtered by the PSF, and thus are
easier to detect.  We estimate the significance of each galaxy according 
to $\nu = f/(\sqrt{n} \sigma)$, 
where $f$ is the weighted flux of the object, 
$n$ is the white noise density of the sky, 
and $\sigma$ is the scale size of the object.  
If these are all measured in the {\em convolved} images, 
which have round PSF,
then the estimate of 
$\nu$ has no shape bias.  We have found empirically that {\tt SExtractor} 
detects essentially all objects with $\nu > 10$, so galaxies more 
significant than this are unaffected by selection bias.  We therefore
cut from the galaxy sample all galaxies with $\nu < 10$.
\end{enumerate}

We now have a catalog of galaxies for each field, 
with an estimate of the true shape of each galaxy before being 
affected by seeing or other PSF effects.  We also keep in this catalog 
an estimate of the errors in the ellipticity measurements, the magnitude,
and the size of the galaxy.

One may ask how our BJ02-based methodology compares to the more common 
KSB methods. We have not tried to construct a parallel KSB pipeline 
for our data, so we can only speculate on the relative merits.
From theoretical considerations (see BJ02), we expect our methods to 
reduce the impact of photon noise on the measurements of individual 
ellipticies by 20\% or so. 
For circular objects the 2 methods are equivalent. 
There is a similar potential S/N gain from the use of our 
nearly-optimal weighting when combining shapes to estimate a distortion 
(\S\ref{determiningshear}), since the shape noise is minimized. 
We expect a more significant benefit over KSB to be
a reduction of systematic errors due to uncorrected PSF ellipticities 
(described in next section).
KSB breaks down when objects aren't Gaussian or the PSF isn't nearly 
circular, and our higher-order circularization kernel is expected 
to do a better job with the $e\sim0.2$ PSFs that are present in our data. 

\subsection {Checks for Systematic Errors}
\label{systematics}

\begin{figure}[t]
\epsscale{0.9}
\plottwo{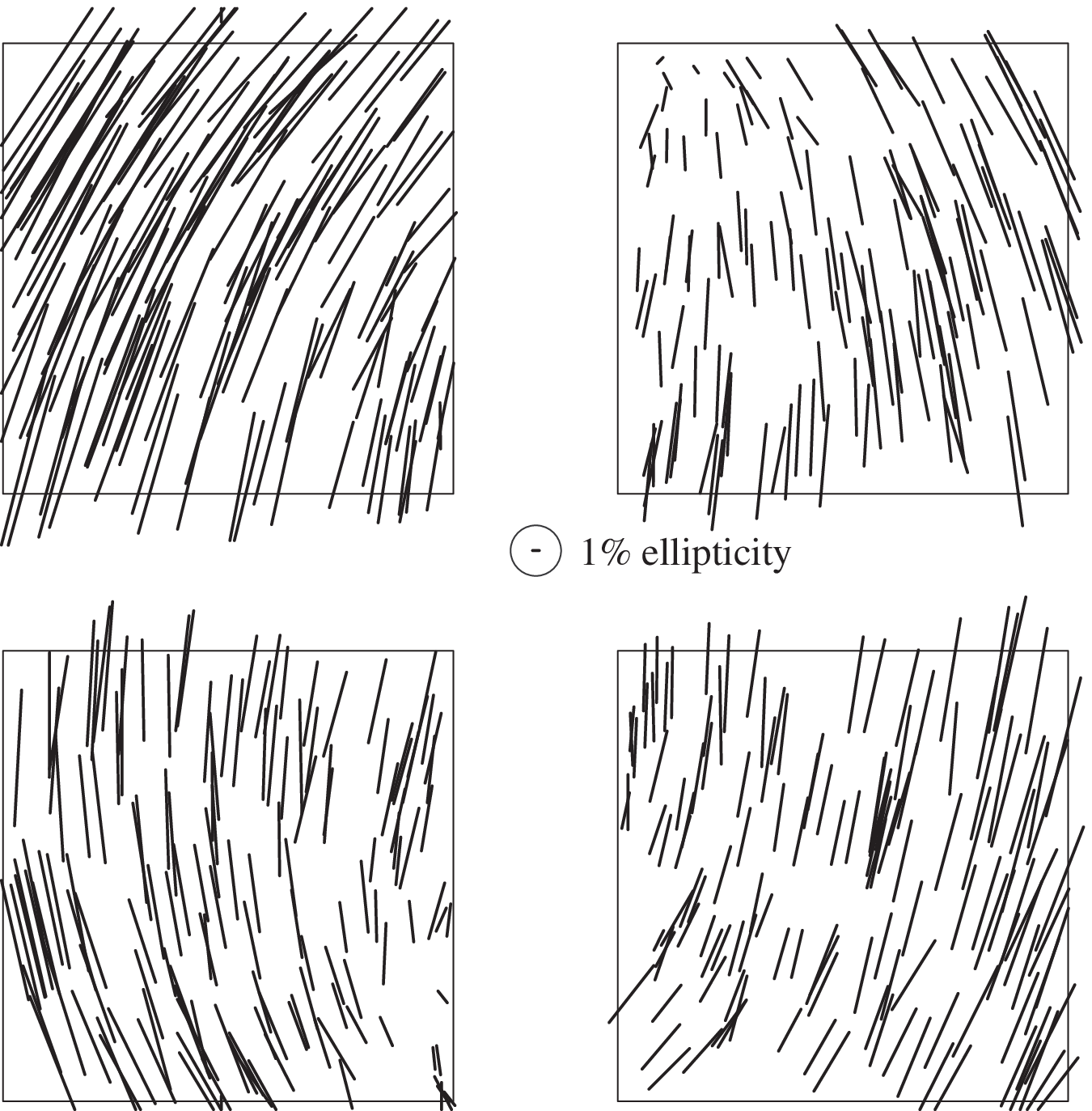}{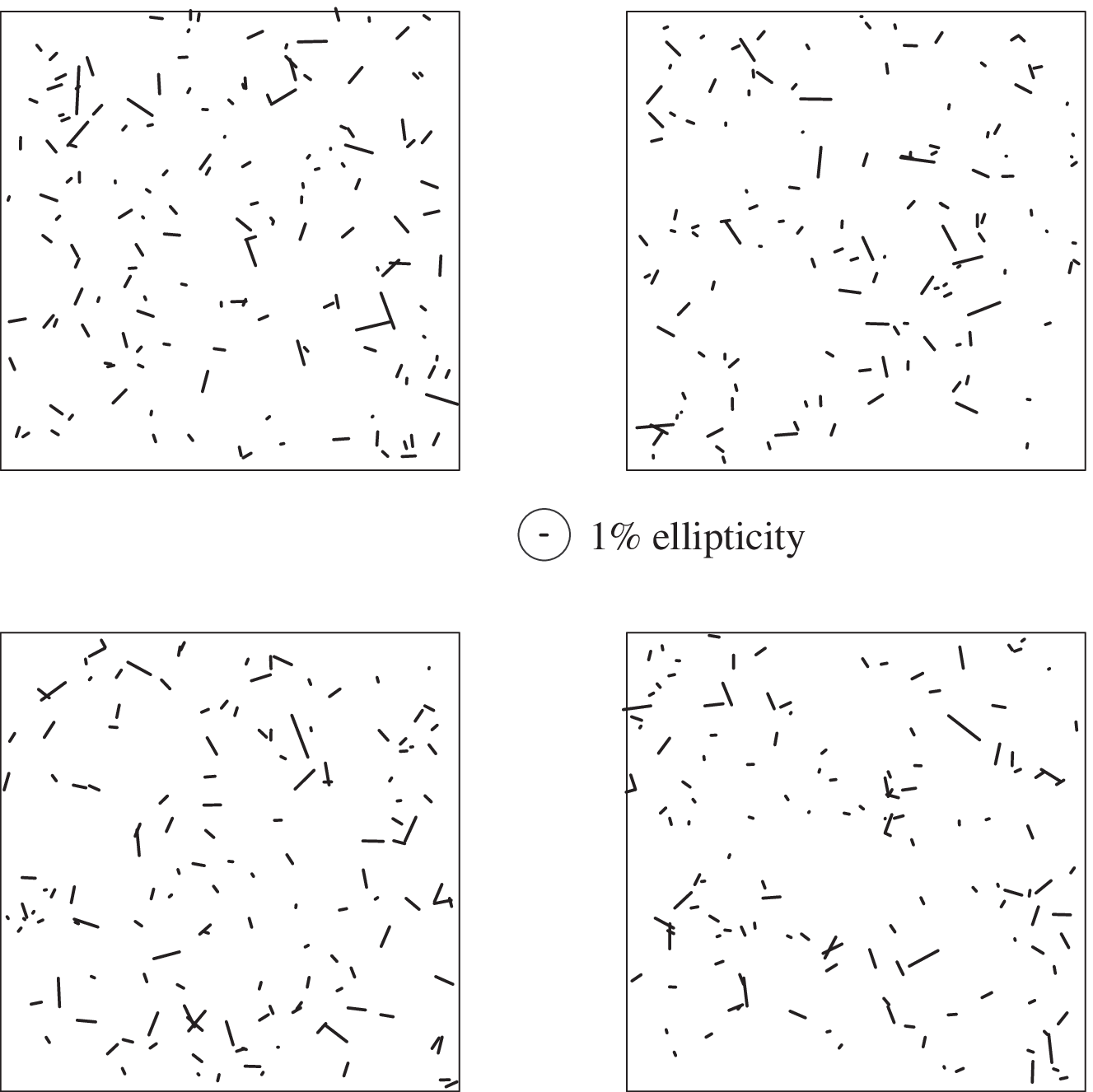}
\caption[]{ \small 
Whisker plots of star ellipticities before and after processing for
one of our BTC exposures.
The length of each ``whisker'' is proportional to the magnitude of the
ellipticity, and the orientation corresponds to the direction of the
ellipticity.  
The whisker in the center corresponds to a 1\% ellipticity.
The remaining ~1-2\% ellipticity values after processing are
seen to be essentially uncorrelated and are primarily
due to measurement noise.  
}
\label{starwhiskerplots}
\end{figure}

The most daunting problem in weak-lensing measurements is not the
volume of the data, but rather the elimination of spurious shape
correlations caused by instrumental effects, particularly asymmetric
PSFs.  Our decision to make a shallow survey is beneficial in that the
galaxies have known $N(z)$ and larger angular sizes than faint
galaxies, but a shallow survey has a smaller lensing signal and hence
is more susceptible to systematic errors.  Our requirements for
rejection of systematic signals are very stringent:  as noted below,
the lensing distortion signal at our largest scales is $\approx0.3\%$ RMS,
while a large fraction of our images contain stellar images with
ellipticities of 10\% or higher.  This is particularly a problem for
the BTC images, because significant astigmatism in the telescope
combines with warped CCDs to produce larger PSF ellipticities than is
typical.  Galaxies measured in image regions with stellar ellipticities
$>0.25$ are discarded from the catalogs.

There are a number of checks one can make to look for systematic errors in 
the galaxy shapes.  In particular, since the convolution should make the 
stars round, we check that the stars in the final image actually 
are round.  Also, it is expected that most systematic 
effects will correlate with the position on the chips and/or the size and
shape of the PSF.  We describe here some of the tests we have done to
quantify systematic errors which might remain after the processing
steps described in \S\ref{datareduction}.

\subsubsection {Final Star Shapes}
\label{finalstarshapes}

We make a ``whisker plot'' of the stars in every image to 
look for images where the processing may have gone wrong.  For example,
missed stars near the corners of the chip, a bad fit to the kernel, 
\etc\ A representative example of one of these plots along with the 
corresponding plot before the convolution is given in 
Figure~\ref{starwhiskerplots}.  For reference,
the horizontal whisker in the center of each corresponds to 1\%
ellipticity. 

Clearly, the convolution does not make each star exactly round.  Each whisker
on the plot still has about 1-3\% ellipticity.  
The important thing to check is that the convolution has removed any
coherence in the whisker orientations.  Measurement error on the
stars' ellipticities will leave uncorrelated random whiskers as
residuals. The 
longer whiskers tend to correspond to fainter, noiser stars.

\begin{figure}[tp]
\epsscale{0.9}
\plottwo{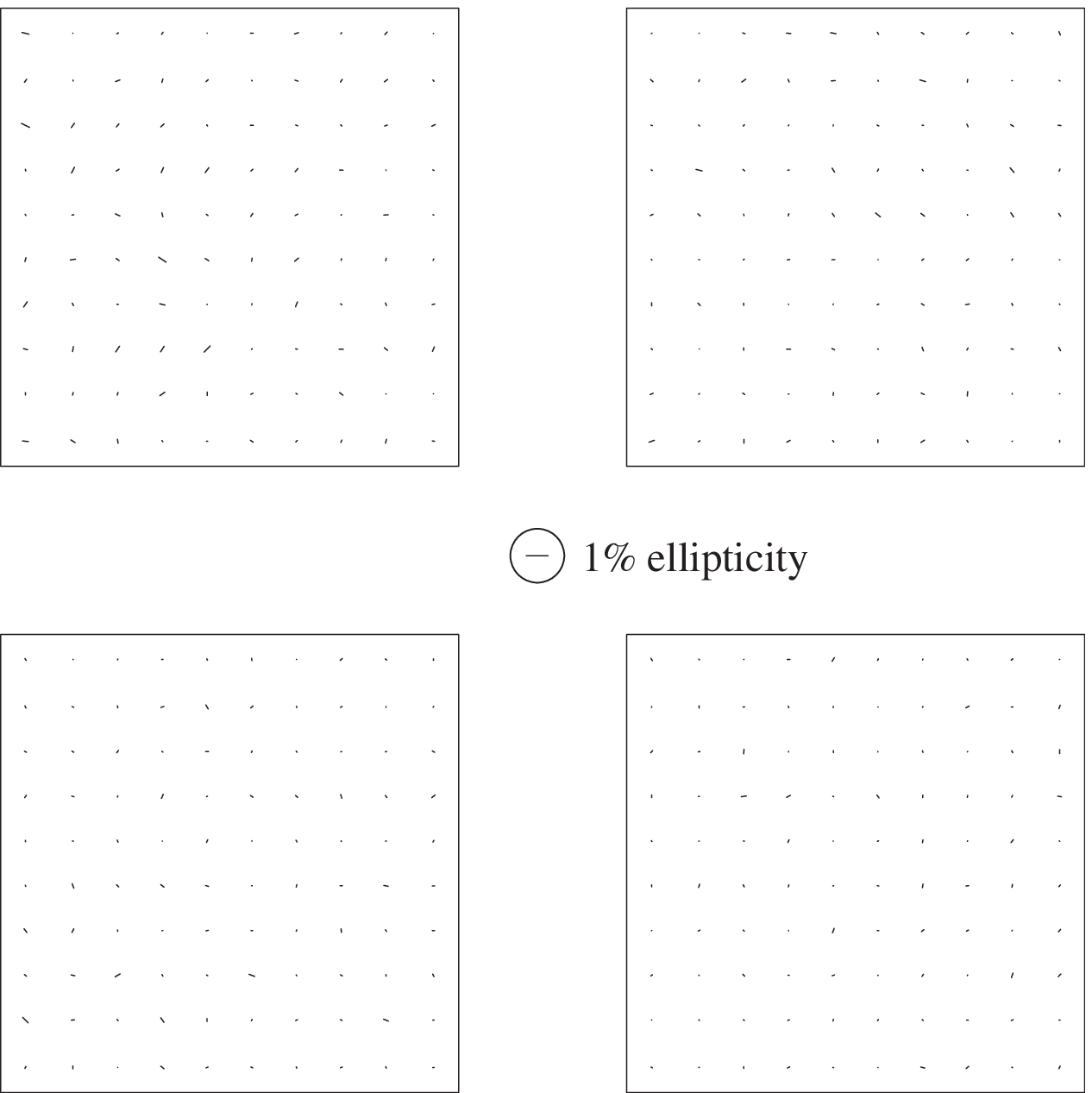}{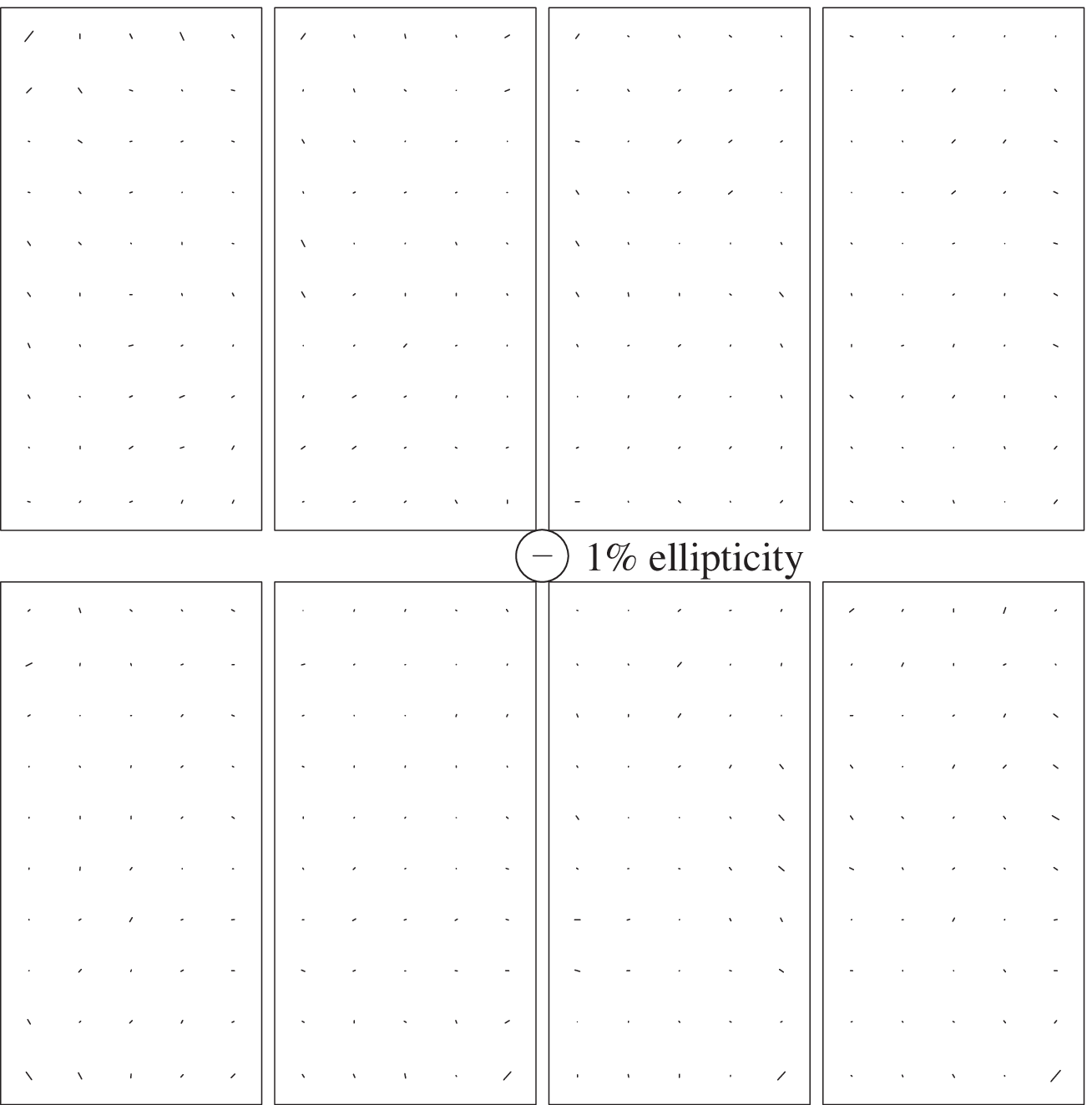}
\caption[]{ \small 
Post-processing star shapes binned according to chip position for each of 
the BTC (left) and Mosaic (right) chips.  
The whiskers indicate the magnitude and orientation of this average shape.
The central whisker in each plot corresponds to 1\% ellipticity.  
The slight residuals which remain are well below the 1\% level.
}
\label{psfshapeplots}
\end{figure}

To search for residuals which may be
coherent functions of pixel coordinates,
we average the shapes of the stars from many images as a function
of position on the CCD array.  The whisker plots resulting from this 
procedure for both the BTC and Mosaic chips
are shown in Figure~\ref{psfshapeplots}.

All of the whiskers are smaller than the 1\% whisker, with 
most of the whiskers barely visible ($<0.1\%$).  The only 
whiskers which approach 1\% in size are in the corners of the 
chips.  This is mostly due to the fact that stars near the corners
are more likely to be rejected from the kernel fits
(in step~\ref{stepconvolve} of \S\ref{datareduction}).  
Therefore, fewer stars are being averaged, and the statistical
errors tend to increase the resultant whisker size.  This is
not a problem for the galaxies, because galaxies in regions with 
no stars are also rejected.

\begin{figure}[tp]
\epsscale{1.0}
\plotone{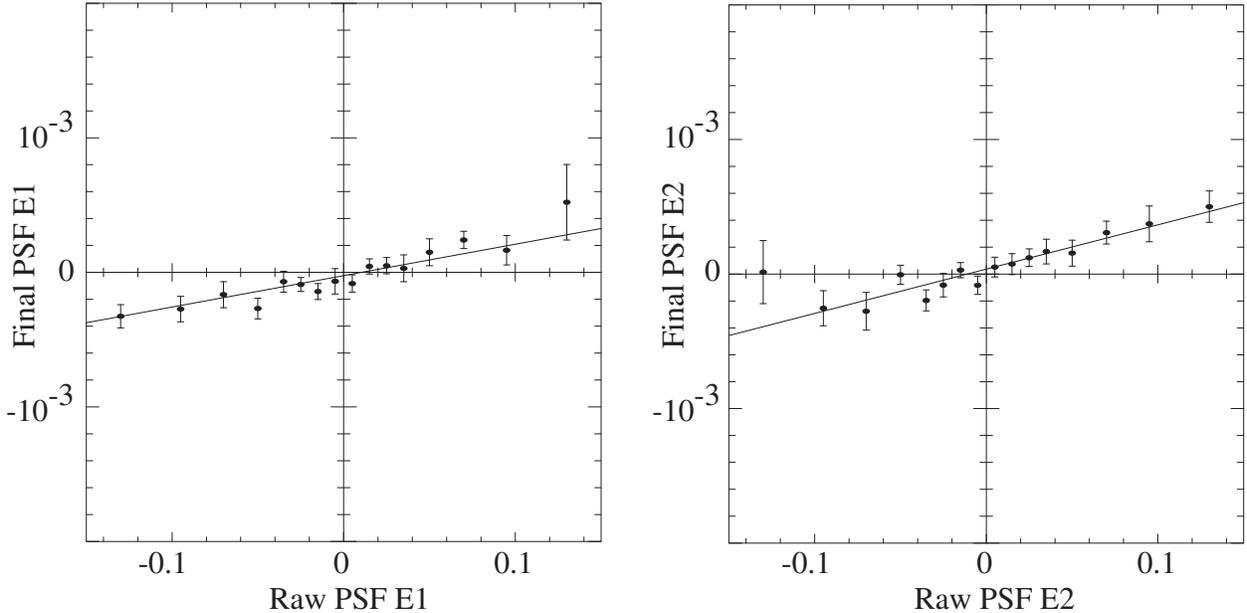}
\caption[]{ \small 
Final post-processing shapes of the stars binned according to their
initial observed shape for each of the two components of the 
ellipticity.  The fitted slope is of order 1/300, leaving an rms 
residual effect of 0.03\% and a maximum effect of less than 0.1\%,
which is well below the strength of the lensing signal.
}
\label{residpsfplot}
\end{figure}

Another way to see how well we are making the stars round is to plot the
final shape of the stars against the initial shape.  This plot is shown in
Figure~\ref{residpsfplot}.  There is a noticeably positive 
slope for both $e_+$ and $e_\times$, but the slope is of order 1/300.
This means that the shape of the PSF is reduced by a factor of 300 by the
convolution.  The mean final ellipticity for the worst initial ellipticities 
is less than 0.1\%, which is well below the level of our lensing signal.  
Moreover, the rms deviation from zero is only 0.03\%.

\subsubsection {Galaxy Shape vs. Chip Position}
\label{galshapevschip}

\begin{figure}[tp]
\epsscale{0.9}
\plottwo{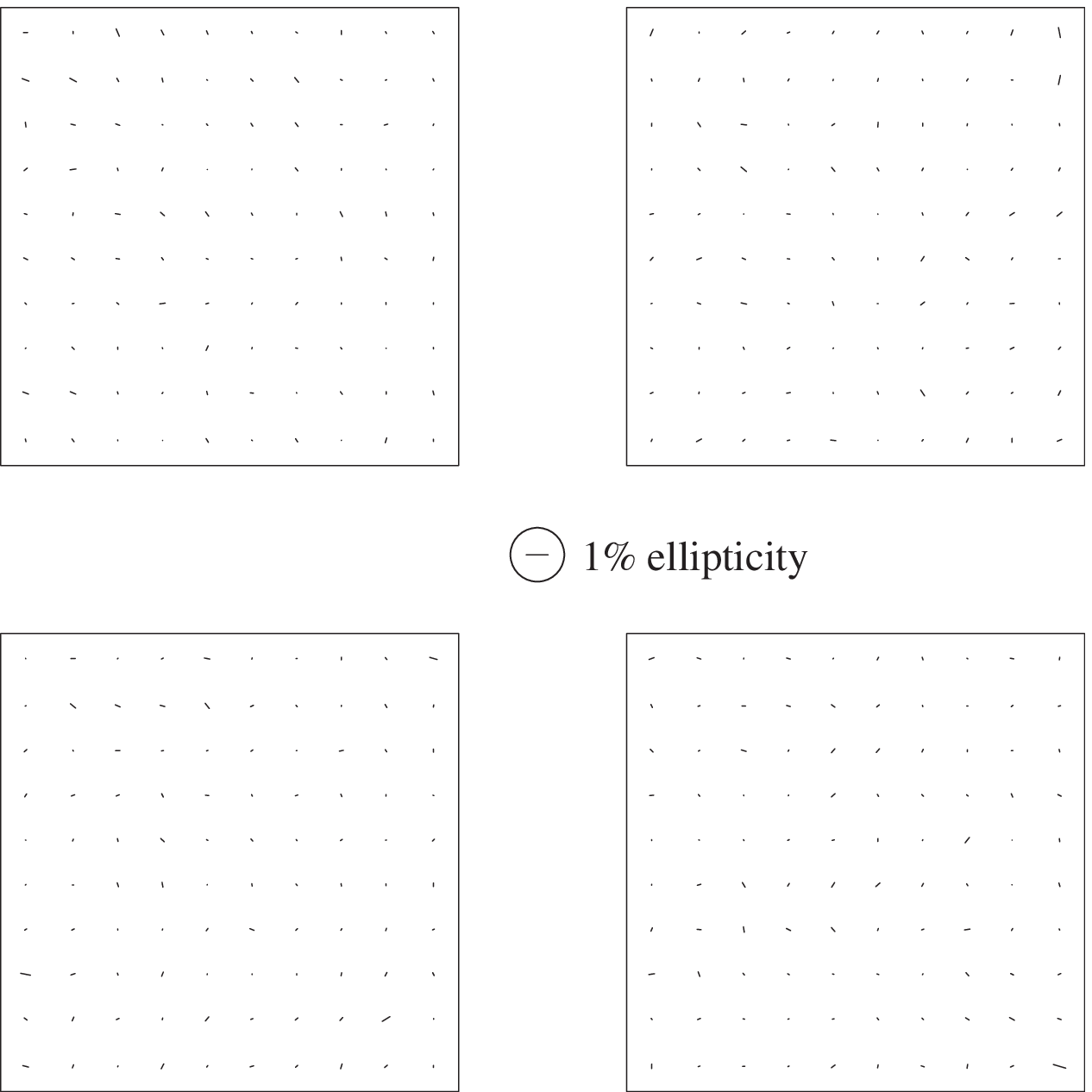}{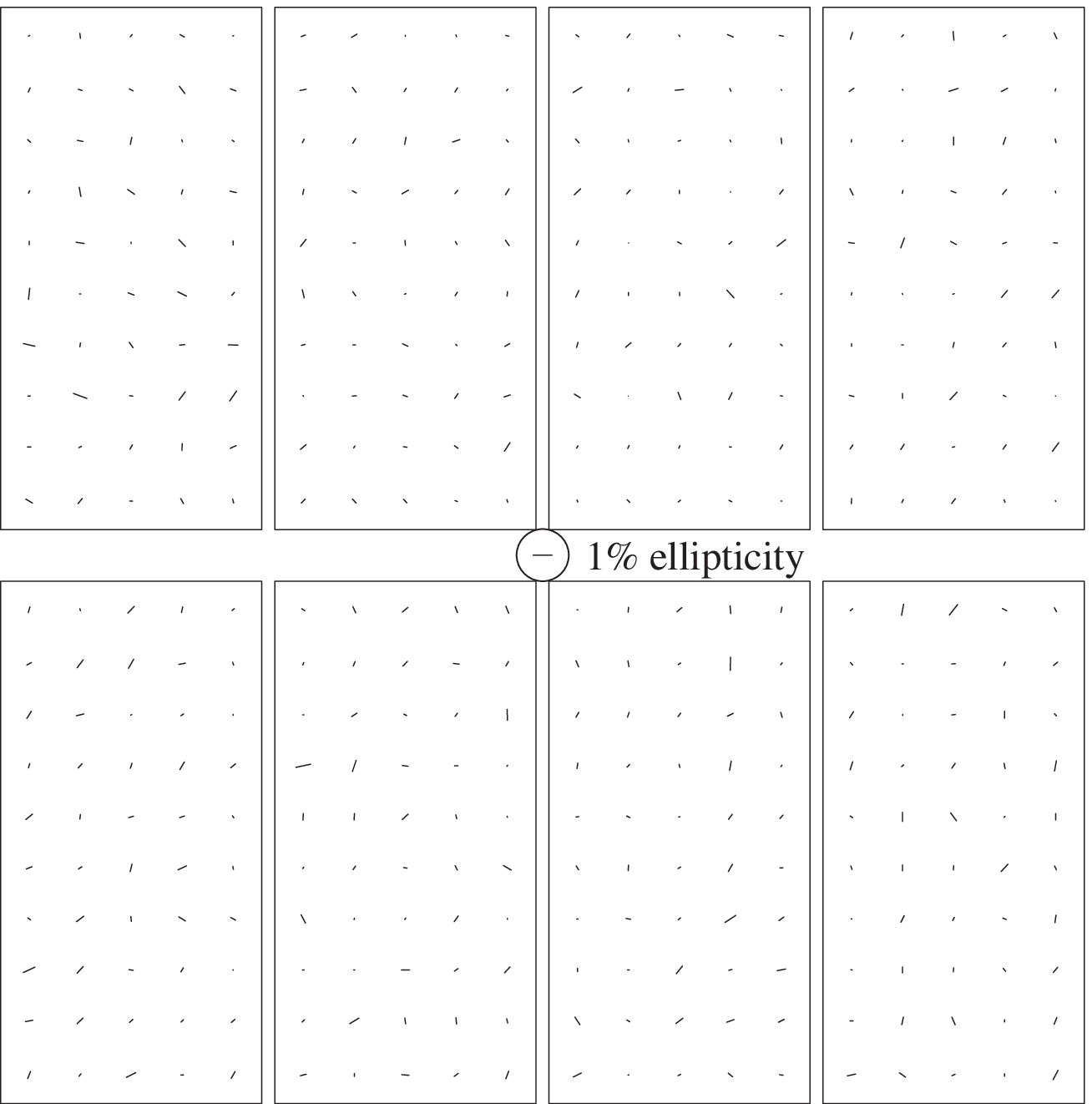}
\caption[]{ \small 
Post-processing galaxy shapes binned according to chip position for 
the BTC (left) and Mosaic (right) chips.  For each shape observation, we 
subtract off the mean shape for that object as calculated
from several observations from widely separated chip positions 
before binning.  This removes the effect of shape noise.  
The central whisker in each plot corresponds to 1\% ellipticity.  
}
\label{galshapeplots}
\end{figure}

We can make whisker plots for the galaxies as well as the stars.
The galaxies, however, are not each expected
to be round after convolving.  Even when the PSF is round, the galaxies
each have a real shape variance of order 30\%.  To see if the
convolution has 
affected the galaxies correctly as well as the stars, we reduce
the shape noise by averaging the shapes of many galaxies.
We first bin the galaxies by array position to search for
systematic errors that depend upon pixel coordinates.
Figure~\ref{galshapeplots} shows the (null) results of this test.
For neither hardware configuration do we detect any residual pattern
in the galaxy shapes.

\subsubsection {Galaxy Shape vs. PSF Shape}
\label{galshapevspsf}

The biggest systematic effect to eliminate is the effect of 
the PSF.  While there are other effects which vary across the chip, such as 
charge transfer inefficiencies, telescope distortion, and the flat field 
pattern, clearly the main concern is the PSF.

\begin{figure}[tp]
\epsscale{1.0}
\plotone{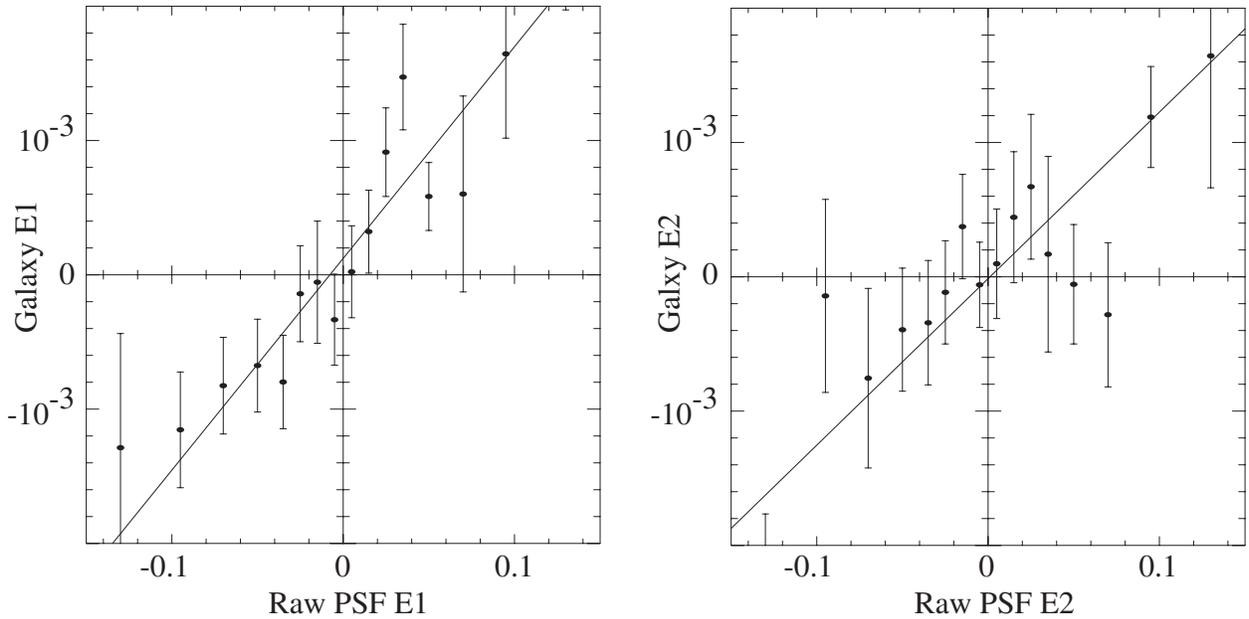}
\caption[]{ \small 
Final post-processing galaxy shapes binned according to the
PSF where they were observed.  
Clearly, there is still some
bias relative to the PSF with a slope of order 0.015, which corresponds
to a maximum bias of 0.4\% for our worst PSFs, and a 0.2\% rms effect.
Note the greatly expanded vertical scale.
}
\label{biasplot}
\end{figure}

The most obvious test then is to bin the allegedly-corrected galaxy shapes
by the ``raw'' PSF shape, as was done for the convolved stars.  
Again, each individual galaxy
is not expected to be round, but the average shape of many galaxies
should be independent of the PSF shapes for the galaxies if the survey
is large enough to decouple the PSF variations from the true lensing
signal. 
Figure~\ref{biasplot} 
shows the results of this test.

Apparently, there is still a bias of the galaxy shapes with respect to the 
initial PSF shape.  
The plot shows
a slope of approximately 0.015, which corresponds to a 0.4\% 
effect for our worst PSFs, and a 0.2\% rms effect.  
This is somewhat below (but of the same order as) the level of
our lensing signal.

We believe the observed residual galaxy dependence upon PSF shape is
primarily
due to higher-order asymmetries in the PSF which have not been removed
by the $7\times7$ convolution filters.  We plan to implement the full
higher-order analytical PSF corrections of BJ02 \S6.3.5 and expect
this to have smaller residuals than the kernel-convolution method used
here.

Until then, we empirically fit for the slope of this bias and 
subtract the bias from the galaxy shape measurements.  
However, it is likely that simply subtracting the bias from the galaxy shapes is
not the correct thing to do.  In particular the bias seems to be
stronger for faint galaxies and for large galaxies (which may
be due to slightly non-linear CCD response or charge transfer
efficiency).  
We currently use different bias slopes for $m > 22$ and $m < 22$, but
it is possible that we have not correctly identified the exact
population of galaxies which are giving us most of the bias.
Therefore, we suspect that this residual is likely the main source of the
B-mode power described in \S\ref{massap}.

\section{Analysis}
\label{analysis}

\subsection{Determining Shear from Shape Averages}
\label{determiningshear}

Once we have a catalog of galaxy shapes, we want to be able to convert these 
shapes into various statistics of the lensing distortion field.  These
statistics
require finding either the average of many ellipticities (\eg
the overall shear in each field, \S\ref{overallshear}) 
or the average product of pairs of ellipticities (\eg the shear
correlation functions, \S\ref{shearcorrelation}).  

The optimal weight for averaging ellipticities to obtain the lensing
distortion [BJ02 Equation (5-28)]
requires knowledge of the intrinsic distribution $P(e)$ of galaxy shapes.  
The distribution for the brighter, well-measured galaxies in this
survey is shown in BJ02 Figure~4.  BJ02  Equation (5-36) gives a 
simple weight which gives nearly optimal results:
\begin{equation}
w = [e^2 + 2.25 \sigma_\eta^2]^{-1/2},
\label{dwts}
\end{equation}
where $\sigma_\eta$ is the measurement uncertainty in each component
of the shape, as measured in the sheared coordinate
system where the shape is circular.  
BJ02 Figure~5 demonstrates that this weight 
recovers very close to the optimal signal to noise for the estimate of the 
distortion.

Thus, our estimate for a distortion from a set of shapes is 
(from BJ02 Equations~5-23, 5-33, and 5-35)
\begin{align}
\label{dsum}
\hat{\bfd} 
&= \frac{1}{\cal R} \frac{\sum w \,\bfe}{\sum w} \\ 
\label{dvar}
{\rm Var}(\hat\delta_i)
&= \frac{1}{{\cal R}^2} \frac{\sum w^2 e_i^2}{\sum w^2} 
\qquad (i\in\{+,\times\}) \\ 
\label{Rsum}
{\cal R} 
&= \frac{\sum \left[ w \left(1 - k_0 - \frac{k_1 e^2}{2} \right)
                   + \frac{e}{2} \frac{dw}{de} \left(1 - k_0 - k_1 e^2 \right)
              \right]
       }{\sum w} \\
k_0 &= (1-f) \sigma^2_{\rm SN} \\
k_1 &= f^2 \\
f &= \frac{\sigma^2_{\rm SN}}{\sigma^2_{\rm SN} + \sigma^2_\eta}
\label{dend}
\end{align}
The responsivity, ${\cal R}$, is similar to the shear polarizability
of the KSB method and describes how our weighted mean 
ellipticity responds to an applied shear.
In the simple case of an unweighted ellipticity
average with unweighted shape measurements, 
${\cal R} = 1-\langle e^2 \rangle = 1-2\sigma^2_{\rm SN}$.
The shape noise, $\sigma^2_{\rm SN}$, is the variance in the intrinsic 
$e_+$ of the galaxies.  For our brighter galaxies, we measure 
$\sigma_{SN}=0.31$.

Figure~5 of BJ02 also demonstrates that the approximations made in the
derivation of the above responsivity lead to $<1\%$ error in the
resultant distortion calibration, for the shapes and noise levels of
galaxies found in this survey.  \citet{Sm01} perform a complete
numerical simulation of the distortion measurement process, concluding
that the overall calibration is accurate to $\lesssim5\%$.  The BJ02
formulae improve upon those of \citet{Sm01} so we believe the
responsivity calibration is now accurate to $\approx2\%$ or better.
 
\subsection{Overall Shear in Each Field}
\label{overallshear}

The simplest statistic to calculate is the average distortion
in each of our  
12 fields. These results are listed in Table~\ref{fieldstable}.  The 
uncertainty on each of these measurements is typically $\pm 1 \times
10^{-3}$.  The $S/N$ for detection of each distortion component in
each field is $\approx 2.5$. Collectively, they give a
strong detection of the rms fluctuation in the shear field averaged
on a scale of $2\fdg5$.  At the $z\approx0.25$ redshift at which our
sensitivity to lensing matter peaks (\cf \S\ref{zdist}), this corresponds to a
comoving smoothing scale of $\approx25h^{-1}$~Mpc, the largest scale
to date on which gravitational lensing effects have been detected.

\begin{figure}[t]
\epsscale{0.7}
\plotone{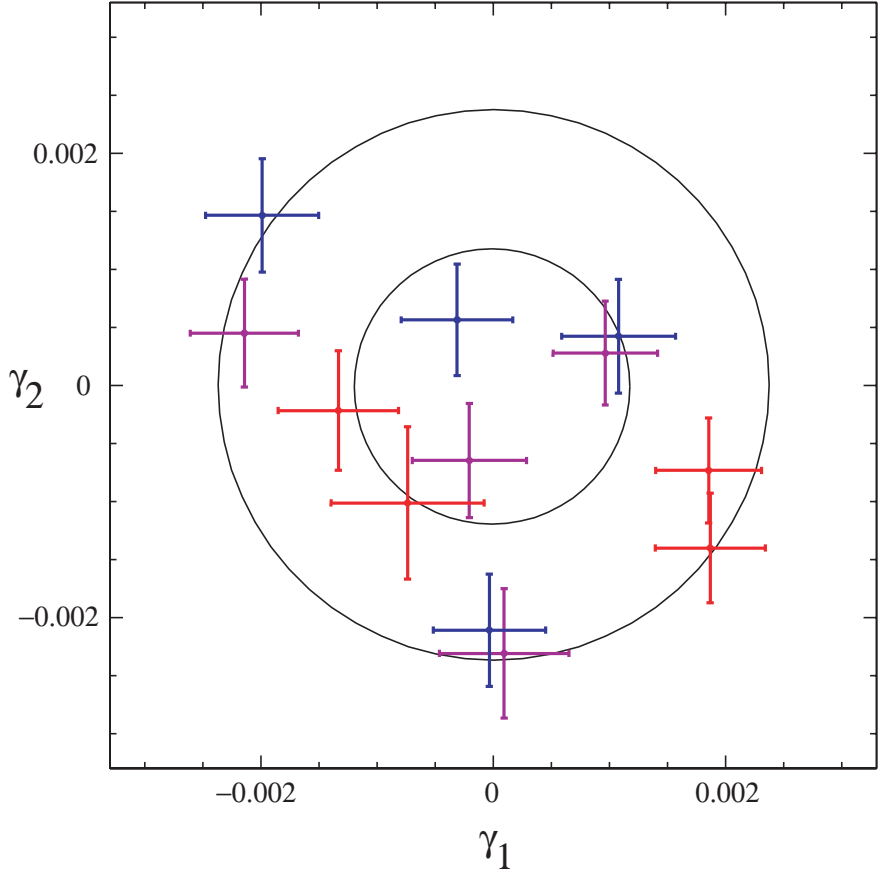}
\caption[]{ \small 
Scatter plot of the mean shear in each of the 12 fields.
The distribution is found to be consistent with a Gaussian
with $\sigma = 0.0024 \pm 0.0006$.  This is in addition
to the (assumed Gaussian) measurement errors.
The red crosses are the Mosaic fields, blue are BTC, and purple
are mixed.
}
\label{scatterplot}
\end{figure}

Figure~\ref{scatterplot} shows a scatter plot of the 12
distortion values, along with their error bars for each component.
The mean distortions appear randomly distributed as expected for a
real signal.  A tendency to align on the $\delta_+$ axis would
indicate a systematic error aligned with the CCD axes, such as charge
transfer nonlinearities---no such effect is seen.

We assume that the average of each distortion component $\delta_+$
or $\delta_\times$ in a 2\fdg5 square box has a
Gaussian probability distribution with width $\sigma$.
The expected distribution is then broadened by the measurement error
in each field (\eqq{dvar}).
A maximum likelihood analysis of the 24 values 
then yields $\sigma = 0.0024 \pm 0.0006$.
Circles at 1$\sigma$ and 2$\sigma$ are also drawn on 
Figure~\ref{scatterplot} for reference.
For comparison to other results which use the shear, $\gamma$,
rather than the distortion, 
$\langle\gamma^2\rangle^{1/2} = \sigma/2 = 0.0012 \pm 0.003$.

\subsection{Shear Correlation Functions}
\label{shearcorrelation}

\citet{Mi91} introduced the shear correlation functions of 
the ellipticities of pairs of galaxies measured with respect to the line
separating them.  We will not make direct use of the correlation
functions to constrain cosmology, but rather follow the prescriptions
of \citet{Cr02} (as clarified in \citet{Pen02} and \citet{Sch02}) 
for constructing the \mapsq\ and \gamsq\ statistics (described
below) from the correlation function data.  

The simplest way to calculate these functions is to treat the lens
distortion as a complex number, $\delta = \delta_+ + i
\delta_\times$.  To conform with other authors, we express the
correlations in terms of the shear $\gamma$, related in the weak limit
to the distortion by $\gamma=\delta/2$.  Three correlation function
are defined by:
\begin{align}
\xi_+(\theta) 
&= \langle \gamma({\bf r}) \gamma^*({\bf r + \bft}) \rangle \\
\xi_-(\theta) + i\xi_\times(\theta) 
&= \langle\gamma({\bf r}) \gamma({\bf r +\bft}) e^{-4i\arg\theta}\rangle
\end{align}
Note that if the galaxies in a pair are swapped, the first equation turns into
its conjugate.  Thus, this quantity can be made manifestly real by double counting
every pair of galaxies.  With only single counting, the imaginary term
gives an estimate of the statistical error of the other three quantities.
Further, taking a mirror image of the entire field will turn the
second equation into its conjugate.  Thus, the imaginary part of this
quantity is also expected to go to zero in the absence of systematic
effects.  

Near-optimal estimators for these two-point functions are constructed
using the weight function and responsivity defined in
Equations~(\ref{dwts}--\ref{dend}):
\begin{align}
\hat \xi_+(\theta) 
&= \frac{1}{4 {\cal R}^2}
   \frac{\sum_{i,j} w_i w_j \delta_i \delta^\ast_j}{\sum_{i,j} w_i w_j} \\
\hat \xi_-(\theta) + i\hat \xi_\times(\theta) 
&= \frac{1}{4 {\cal R}^2}
  \frac{ \sum_{i,j} w_i w_j \delta_i \delta_j e^{-4 i \arg({\bf r}_i-{\bf r}_j)}
      }{ \sum_{i,j} w_i w_j } 
\end{align}
where the sum is taken over all pairs of galaxies $i$ and $j$ with
separation $\theta = |{\bf r}_i-{\bf r}_j|$ within some bin.
The variance of $\hat\xi_i$ due to shape noise and shot noise is
also simply estimated, but cosmic variance and bin covariances are
more difficult to estimate.  We bypass these complications by
constructing field-to-field covariance matrices, as described below.

As our primary use of the correlation functions will be to calculate
other quantities by integrating over a range of $\theta$,
we calculate the correlation functions in fairly small bins with
$\delta(\ln\theta) = 0.05$.  

\subsection{Aperture Mass Statistic}
\label{massap}

The aperture mass statistic (\map) is useful
for estimating cosmological parameters.  The idea 
is fairly straightforward:  a mass concentration at the center of a
given aperture will tend to produce a tangential pattern to the galaxy
ellipticities around the center of the aperture.  For cosmic shear
measurements, we do not detect the mass concentrations individually
[see \citet{Wi01} for a rare (so far) example],
but rather the
magnitude of the mass {\em fluctuations}, so we really want the rms
variation of the \map\ value as the aperture is swept across the sky.

The \map\ statistic for a single aperture of radius $\theta$ is \citep{Sch98}
\begin{align}
\label{mapdef}
M_{ap}(\theta) 
&= \int_0^\theta d^2 {\bf \phi} \,Q(\phi)\, \gamma_t({\bf \phi}) \\
Q(\phi) 
&= \frac{6}{\pi \theta^2} \frac{\phi^2}{\theta^2} 
   \left( 1 - \frac{\phi^2}{\theta^2} \right)
\end{align}
where $\gamma_t$ is the tangential component of the shear 
(as estimated from the galaxy ellipticities).  For many independent 
apertures, $\langle\map\rangle=0$ and the variance,
\mapsq,  probes the power spectrum of the effective convergence.
The window function for this statistic is narrow in $k$-space and
centered at $k\approx 4.1/\theta$ \citep{Sch98}. 

One advantage of the \map\ statistic is that there is a natural test for 
systematics.  If each galaxy is rotated in place by 45 degrees, the
\map\ integral should vanish if due purely to lensing.
This test is essentially measuring the 
curl of the shear field, and is therefore often called the ``B mode,''
while \map\ measures ``E-mode'' power.
We designate this B-mode version of the \map\ statistic as \mx.  Most
systematics are expected to add equal power to the E and B modes,
hence the \mx\ data are
a sensitive test for contaminating systematics.

Another advantage to using the \map\ statistic is that $M_{\rm
ap}(\theta_1)$ 
is very weakly correlated with $M_{\rm
ap}(\theta_2)$ 
when $\theta_1$ differs from $\theta_2$ by a factor of $\approx 2$
\citep{Sch02b}.
Thus, for our range of $1\arcmin<\theta<100\arcmin$, we have
essentially 7 independent points with which to constrain cosmology.  

\begin{figure}[p]
\epsscale{0.63}
\plotone{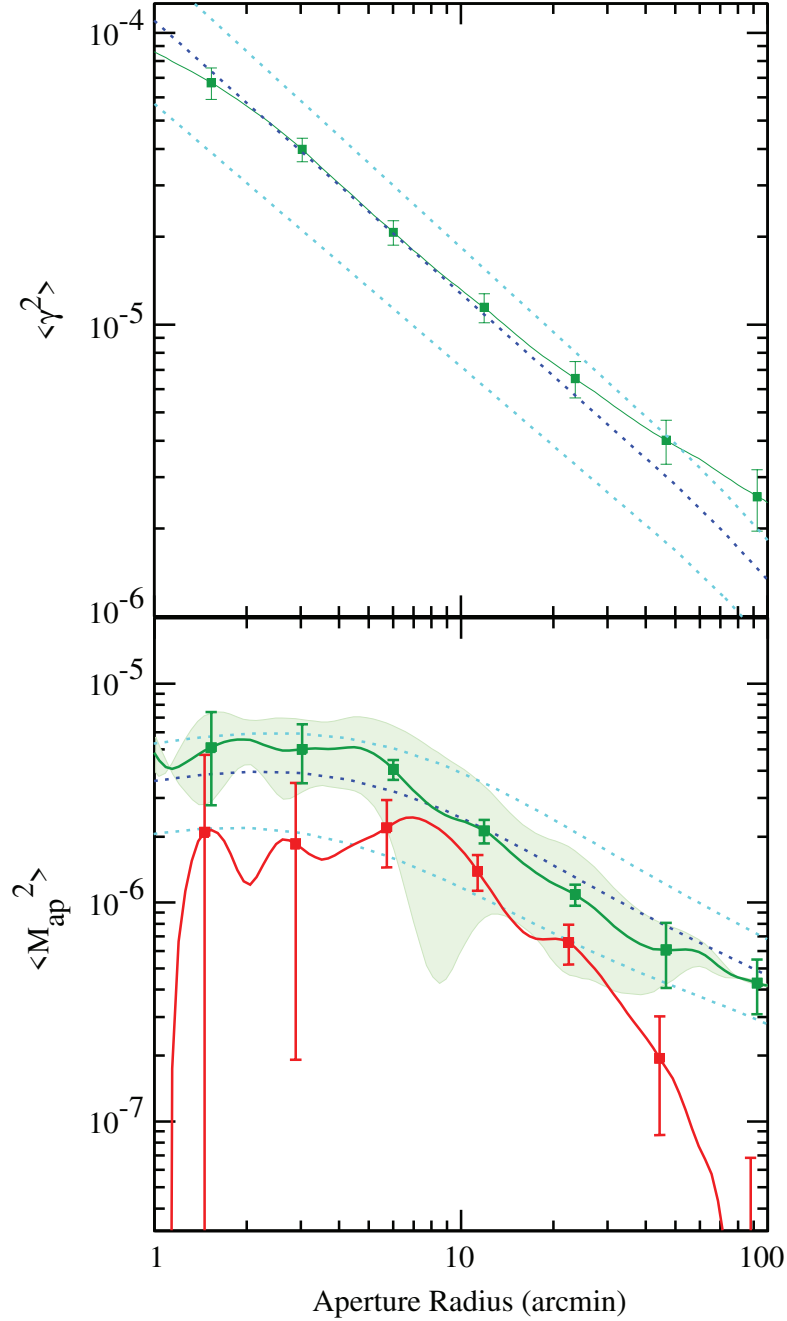}
\caption[]{ \small 
Aperture mass and shear variance statistics measured on 
angular scales from 1 to 100 arcminutes.  For the aperture mass
statistic, the upper green line is the E-mode (lensing) signal, 
and the lower red line is the B-mode (non-lensing) signal.
The light shaded region is bounded by E+B and E-B curves, representing the
range of systematic uncertainty.
Also plotted are the theory curves for our best fit model
(dotted, dark blue), and for models at the 
high and low extremes of our 95\% confidence limits quoted
in \protect\eqq{answer} (dotted, light blue).
Points with error bars are given every 
factor of 2 in radius.  For the \mapsq\ statistic, this 
is approximately the separation at which the data points are
independent of each other.
}
\label{mapgammaplot}
\end{figure}

The problem with calculating \mapsq\ in the obvious way
(scanning the aperture across the images and calculating variance)
is that each aperture is not uniformly filled with galaxies.  
There are holes
due to foreground bright stars, edge effects, bad columns, 
\etc\ These holes
can then bias the resulting \mapsq\ estimates and produce spurious \mxsq\
power.  Specifying a mask for our entire survey would require a
painfully long time, as would the development of software to automate
the task.  
Fortunately, \citet{Cr02} \citep[detailed also in][]{Pen02,Sch02} 
express \mapsq\ and  \mxsq\ as integrals
over the shear correlation functions, which
do not require knowledge of the survey geometry.
The relevant formulae are:

\begin{align}
\langle M_{\rm ap}^2(\theta)\rangle 
&= \frac{1}{2} \int_0^{2\theta} \frac{\phi \,d\phi}{\theta^2}
   \left[ \xi_+(\phi) T_+\left(\frac{\phi}{\theta}\right) 
        + \xi_-(\phi) T_-\left(\frac{\phi}{\theta}\right) 
   \right] \\
\langle M_{\times}^2 (\theta) \rangle 
&= \frac{1}{2} \int_0^{2\theta} \frac{\phi \, d\phi}{\theta^2}
   \left[ \xi_+(\phi) T_+\left(\frac{\phi}{\theta}\right) 
        - \xi_-(\phi) T_-\left(\frac{\phi}{\theta}\right) 
   \right] \\
\end{align}
where equations for $T_+$ and $T_-$ are given in \citet{Sch02}.
The result of this calculation for our data are shown in
Figure~\ref{mapgammaplot} and discussed in the following sections.

Another common statistic of the shear field is the windowed variance,
$\langle\gamma^2(\theta)\rangle$, which is the variance of the shear
when smoothed with a circular window of radius $\theta$.  We have
presented in \S\ref{overallshear} the results for the windowed shear
in 2\fdg5 squares.  For quantitative comparison to cosmological
models, the naive $\langle\gamma^2(\theta)\rangle$ summation would
require knowledge of the survey mask geometry.   Again the above
references show how to express $\langle\gamma^2(\theta)\rangle$ as an
integral over $\hat\xi_+(\theta)$ and $\hat\xi_-(\theta)$, thereby
removing the need to know the mask.  If correlation function data are
available out to separation $\theta_{\rm max}$, then all of \mapsq,
\mxsq\ and \gamsq\ are unambiguously determined for
$\theta\le\theta_{\rm max}/2$.

The \gamsq\ statistic is inferior to \mapsq\
in several respects:  its window function in
$k$-space is much broader, and hence measurements at different
$\theta$ are more highly correlated.  Also 
\gamsq\ does not separate E-mode from B-mode
power, so the noise is $\sqrt 2$ higher and there is no
systematic-power null test.  The information in this statistic is
mostly degenerate with the 
\mapsq\ statistic, with the important exception that 
the variance probes the power spectrum for
$k\lesssim 2/\theta$, whereas \mapsq\ probes $k\approx 4/\theta$.
Therefore \gamsq\ for $\theta$ near half our 2\fdg5 field size
probes a portion of the power spectrum at larger physical scales than
is accessible to \mapsq.   We will therefore make use of \gamsq\ for
$\theta>50\arcmin$. 
The \gamsq\ statistic is plotted above the \mapsq\ statistic in
Figure~\ref{mapgammaplot}. 

\citet{Pen02} and \citet{Sch02} do give prescriptions for producing versions 
of the $\xi$ and
\gamsq\ statistics which separate the E and B contributions.  
However, these expressions have indeterminate constants of integrations
which render the E and B modes degenerate on the scale of the field of
view.  Shear that is nearly constant over the field obviously cannot
be classified as having either E or B properties.  Since we only use
\gamsq\ for values of $\theta$ near the field size, we will not use
this decomposition.

\subsection{Covariance Matrix}
\label{covarmatrix}

For each of our twelve fields, we have a vector of observations,
\({\bf x}_n\ (n=1,\ldots,12)\),
including the \mapsq\ values from 1\arcmin\ to 100\arcmin, and the 
\gamsq\ values from 50\arcmin\ to 100\arcmin.  
Each of these vectors has many data points, since we calculate
each statistic at rather small intervals in aperture radius
$\theta$.  However,
these data are highly degenerate in their information content.
\mapsq\ values become essentially independent at a factor of
2 in $\theta$, giving only about 7 independent data points.  
And \gamsq\ values are highly correlated with each other, 
with the effect that these values only add one more 
independent data point.  So the mean of the twelve vectors,
${\bf x} = \langle{\bf x}_n\rangle$, gives 
us essentially 8 independent points with which to constrain
cosmology.

To quantify this degeneracy more exactly, we construct the full
covariance matrix, $\Sigma$, for the $N=12$ vectors by equally
weighting each of the twelve fields:
\begin{equation}
\Sigma_{ij} 
= \frac{ \sum_n [({\bf x}_n)_i - \langle x_i \rangle] 
                [({\bf x}_n)_j - \langle x_j \rangle]
      }{N(N-1)}
\end{equation}
where $n$ ranges over the $N=12$ fields, and $i,j$ are the indices
of the data values.

This construction of the covariance matrix means that we
ignore the (approximately
identical) nominal error bars for each point, using the 
actual field-to-field variation as our estimate of the error.
This has the advantage that it automatically includes 
cosmic variance in the uncertainty as well as measurement noise
both on and off the diagonal.  
The disadvantage is that
the half-size fields, which have slightly larger statistical
error bars, are given equal weight to the full-size fields.

\subsection{B-mode Power}
\label{bmode}

It is evident in Figure~\ref{mapgammaplot} that there is
significant power in the \mxsq\ statistic, indicating that 
we do have some B-mode power, and hence some systematic 
contamination in our data.  The effect becomes much weaker
at scales $\theta > 30\arcmin$, suggesting that the 
large scale data are probably free of this contamination.
Since the B-mode drops to essentially zero at the 
largest scales of \map, we also expect that the 
\gamsq\ statistic will be free of the contamination
for $\theta>50\arcmin$, since it probes power at even larger angular
scales. 

Note that the observed B-mode signals are much larger than those
to be expected from intrinsic galaxy-shape correlations \citep{Cr02} or
second-order gravitational lensing effects \citep{Sch02,Coo02}.
We believe that they are more likely due to uncorrected high-order
PSF effects or inexact kernel fitting, 
both of which become more important at smaller scales.

Most systematics effects, including uncorrected PSF variation,
can increase \mapsq\ and \mxsq\ 
much more easily than decrease.  However, one can also conceive
of systematic effects which simply mix
power from the E-mode into B-mode rather than add power to either one.  
For example, if each 
ellipticity vector's orientation is rotated slightly but the
magnitude is unchanged, then no power is added, but the power
is mixed somewhat into the B-mode.  

Therefore, a conservative estimate is that the \mapsq\ values
can be in error in {\em either} direction by the amount of the \mxsq\ 
values.  Further, it is not sufficient to simply increase the
error bars on individual data points by this amount, 
since the effect is presumably in 
the same direction at all (or many) values of $\theta$.  Instead,
one should consider the two cases of adding the B-mode to all points
and of subtracting it from all points.  The range
of these two cases will then give an estimate of the full potential
effect of the systematic error.  If the spurious systematic signal has
more E power than B power, we could still have overestimated the
lensing E signal.  Such behavior is expected only, however, for
intrinsic correlations, which we believe to be a small constituent of
our systematic contamination.

In any case, for the purpose of constraining cosmology, 
we want to limit our consideration to the range of $\theta$
that has the 
least amount of B-mode contamination. We will use the
\mapsq\ values only for $30\arcmin < \theta < 100\arcmin$, and still
use \gamsq\ from $50\arcmin < \theta < 100\arcmin$.  
Thus, our data vectors ${\bf x}_k$ (\cf \S\ref{covarmatrix}) 
now only have essentially 4 independent data points.  
The increase in the statistical error from this reduced $\theta$ range is
however much more than compensated by the decrease in the systematic error
from the B-mode; in this range the systematic error is smaller
than the statistical error.

\begin{deluxetable}{rcccccccccc}
\tablewidth{0pt}
\tablecaption{Data Used for Cosmological Constraints}
\tablehead{
\colhead{} & 
\colhead{E-Mode} &
\colhead{\(\sigma_E\)} &
\colhead{B-Mode} &
\colhead{\(\sigma_B\)} &
\multicolumn{6}{c}{Reduced Covariance Matrix} \\
\colhead{} &
\colhead{$\times 10^{-7}$} &
\colhead{} &
\colhead{$\times 10^{-7}$} &
\colhead{} &
\colhead{} &
\multicolumn{4}{c}{($\Sigma_{ij}/\sigma_i\sigma_j$)} &
\colhead{} 
}
\startdata
$\mapsq(30\arcmin)$ & 9.52 & 1.38 & 3.44 & 1.25 &
  & 1 & 0.80 & 0.20 & -0.13 & \\
$\mapsq(50\arcmin)$ & 7.24 & 2.00 & 1.23 & 1.19 &
  & 0.80 & 1 & 0.14 & 0.02 & \\
$\mapsq(100\arcmin)$ & 3.84 & 1.09 & -0.21 & 1.00 &
  & 0.20 & 0.14 & 1 & -0.11 & \\
$\gamsq(100\arcmin)$ & 26.4 & 6.20 & - & - &
  & -0.13 & 0.02 & -0.11 & 1 &
\enddata
\label{datatable}
\end{deluxetable}

\begin{figure}[t]
\epsscale{0.8}
\plotone{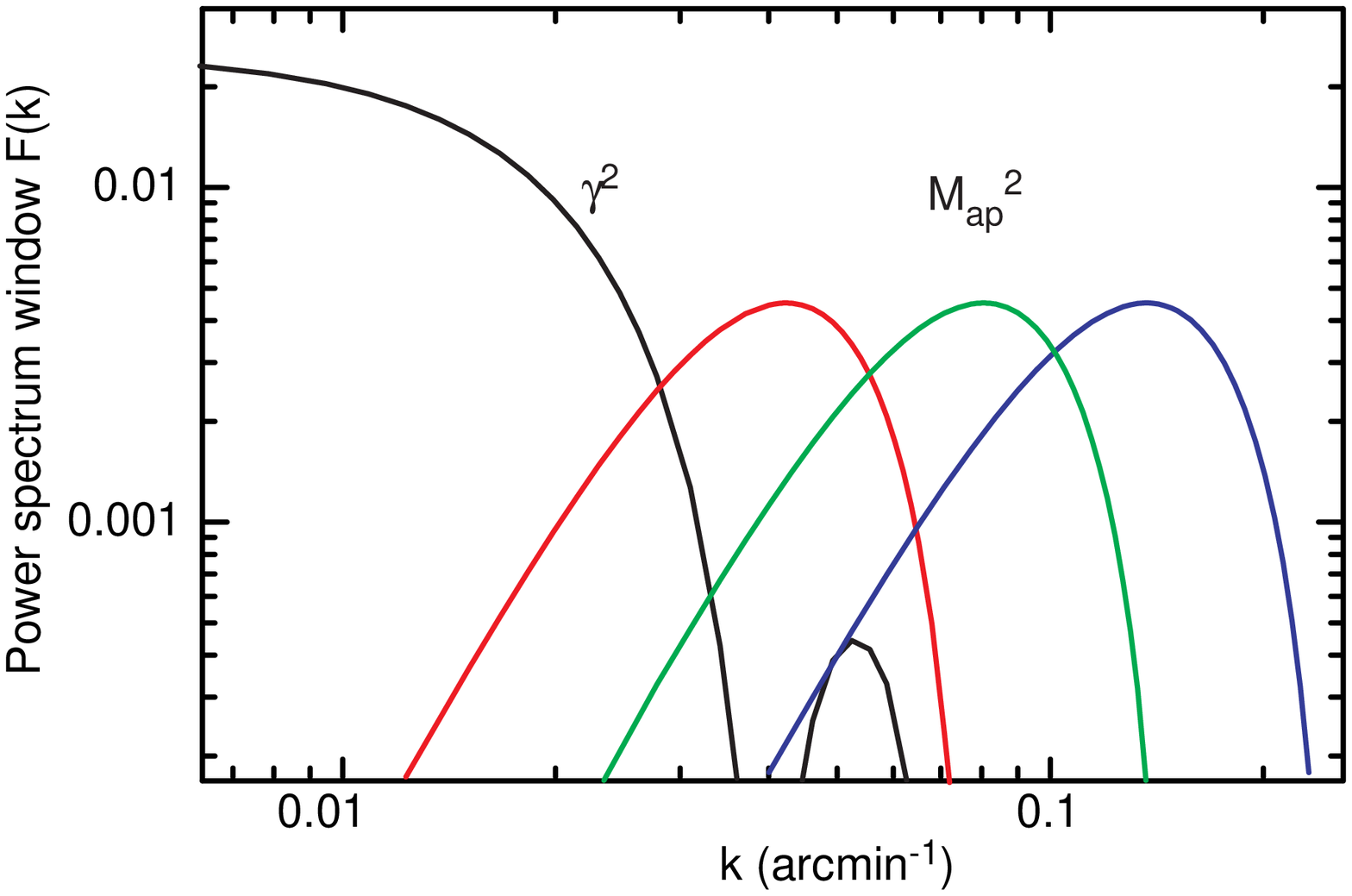}
\caption[]{ \small 
Window functions for each of the statistics listed in 
Table~\protect\ref{datatable}, as defined in \eqq{filterdef}.  
From left to right, the curves are for $\gamsq(100\arcmin)$ (black),
$\mapsq(100\arcmin)$ (red), $\mapsq(50\arcmin)$ (green), 
and $\mapsq(30\arcmin)$ (blue).
}
\label{filtersplot}
\end{figure}

For reference, an abridged listing of the data and covariance matrix
is given in Table~\ref{datatable}.  It includes both the shear
variance and the  
aperture mass statistics (both E and B mode in the latter case)
for selected values within this range.  It also gives the reduced
covariance matrix as described in \S\ref{covarmatrix} for these values.
It is evident that the four statistics are relatively uncorrelated
with only $\mapsq(30\arcmin)$ and $\mapsq(50\arcmin)$ being somewhat
correlated.\footnote{
\citet{Ta02} predict that the kurtosis of the shear field
is beginning to be significant
on these scales and could be the cause of this correlation.}
This is expected, since they sample the power spectrum
at relatively disjoint ranges in frequency.  Each statistic can
be written:
\begin{equation}
y_i = \int k dk P_\kappa(k) F_i(k\theta)
\label{filterdef}
\end{equation}
and Figure~\ref{filtersplot} shows $F_i$ for each of these four statistics.

\section{Cosmological Implications}
\label{cosmo}

\subsection{Redshift Distribution}
\label{zdist}

In BJ02 it is shown that the distortion estimators in
\S\ref{determiningshear} will converge to the lensing distortion.  In
reality the expected distortion $\delta(z)$ is a function of the
source redshift $z$, and we determine some mean distortion
$\bar\delta$.  We do not know the redshift $z_i$ of every source
galaxy, but we can have some knowledge of its distribution vs.
magnitude $P(z|m)$ from spectroscopic redshift surveys.  If we divide
our source galaxies into redshift bins $Z_j$ then the measured signal
should converge to
\begin{align}
\bar\delta = \langle\hat\delta\rangle 
&= \frac{\sum_i w_i e_i}{\sum_i w_i {\cal R}_i} \\
&= \frac{\sum_j \left[ 
           \left( \sum_{z_i\in Z_j} w_i \right) 
           \frac{ \sum_{z_i\in Z_j} w_i {\cal R}_i}{\sum_{z_i\in Z_j} w_i}
           \frac{ \sum_{z_i\in Z_j} w_i e_i}{\sum_{z_i\in Z_j} w_i{\cal R}_i}
          \right] 
       }{\left( \sum_i w_i \right) 
         \frac{ \sum_i w_i {\cal R}_i}{\sum_i w_i} } \\
&= \frac{\sum_j w(Z_j) {\cal R}(Z_j) \delta(Z_j)
       }{\sum_j w(Z_j) {\cal R}(Z_j) }.
\label{rzbins}
\end{align}
Here we have defined $w(Z_j)$ to be the total weight in the redshift
bin, and ${\cal R}(Z_j)$ is the mean responsivity, as per \eqq{Rsum},
of the galaxies in the bin.  We will make the assumption that ${\cal
R}$ is independent of redshift.  For galaxies with magnitude $R<21$,
we do see a significant dependence of the effective ${\cal R}$ 
upon surface brightness, with
$0.62>{\cal R}>0.75$ as we select subsets of extremely high or low
surface brightness and high or low measurement noise.  On the other
hand, each redshift bin contains galaxies with a wide range of surface
brightness and measurement noise, so the responsivity variation with $z$
should be much smaller than this extreme $\pm10\%$ range.

With the ${\cal R}$ taken as constant, the measured distortion
(\ref{rzbins}) becomes
\begin{align}
\bar\delta  
&= \frac{\sum_i w_i \delta(z_i)}{\sum_i w_i } \\
&= \frac{\int\! dz\int\! dw\int\! dm\,\delta(z)  w  P(z,w,m)
       }{\int\! dz\int\! dw\int\! dm\,  w  P(z,w,m)}
\end{align}
where $P(z,w,m)$ is the probability of a given galaxy having redshift
$z$, weight $w$, and apparent magnitude $m$.  Ideally we would
determine the function $P$ by conducting a redshift survey over a
statistically significant sample of galaxies with known $m$ and $w$
for our survey conditions.  Existing surveys, however, can only give
us the conditional distribution $P(z|m)$ of redshift for a given
magnitude.  We will therefore make the further assumption that, within
a given magnitude bin, the redshift $z$ [or more precisely
$\delta(z)$] is statistically independent of the weight $w$, so that
$\langle \delta(z) w P(z,w|m) \rangle = \langle \delta(z) P(z|m)
\rangle \langle w P(w|m) \rangle$, in which case

\begin{equation}
\bar\delta 
= \frac{ \int\!dm \,P(m) 
         \left[\int\!dz \, \delta(z) P(z|m) \right]
         \left[\int\!dw \,w P(w|m)\right]
      }{ \int\! dm \,P(m) \int\! dw\,  w  P(w|m) }
\end{equation}
Since the redshift data are sparse, the integrals are calculated in
0.5-magnitude bins.  Let the lensing survey
field contain a total weight $W(m)$ of galaxies in the magnitude bin,
which we apportion among the $N(m)$ galaxies in the redshift survey
that lie within this bin.  If redshift-survey galaxy $i$ has redshift
$z_i$ and lies in magnitude bin $m_i$, then the
expected signal becomes
\begin{align}
\label{zsum}
\bar\delta &= \sum_i \delta(z_i) W_i \\
W_i &\equiv \frac{W(m_i) / N(m_i)}{\sum_i W(m_i) / N(m_i)}
\end{align}

\begin{figure}[p]
\epsscale{1.0}
\plotone{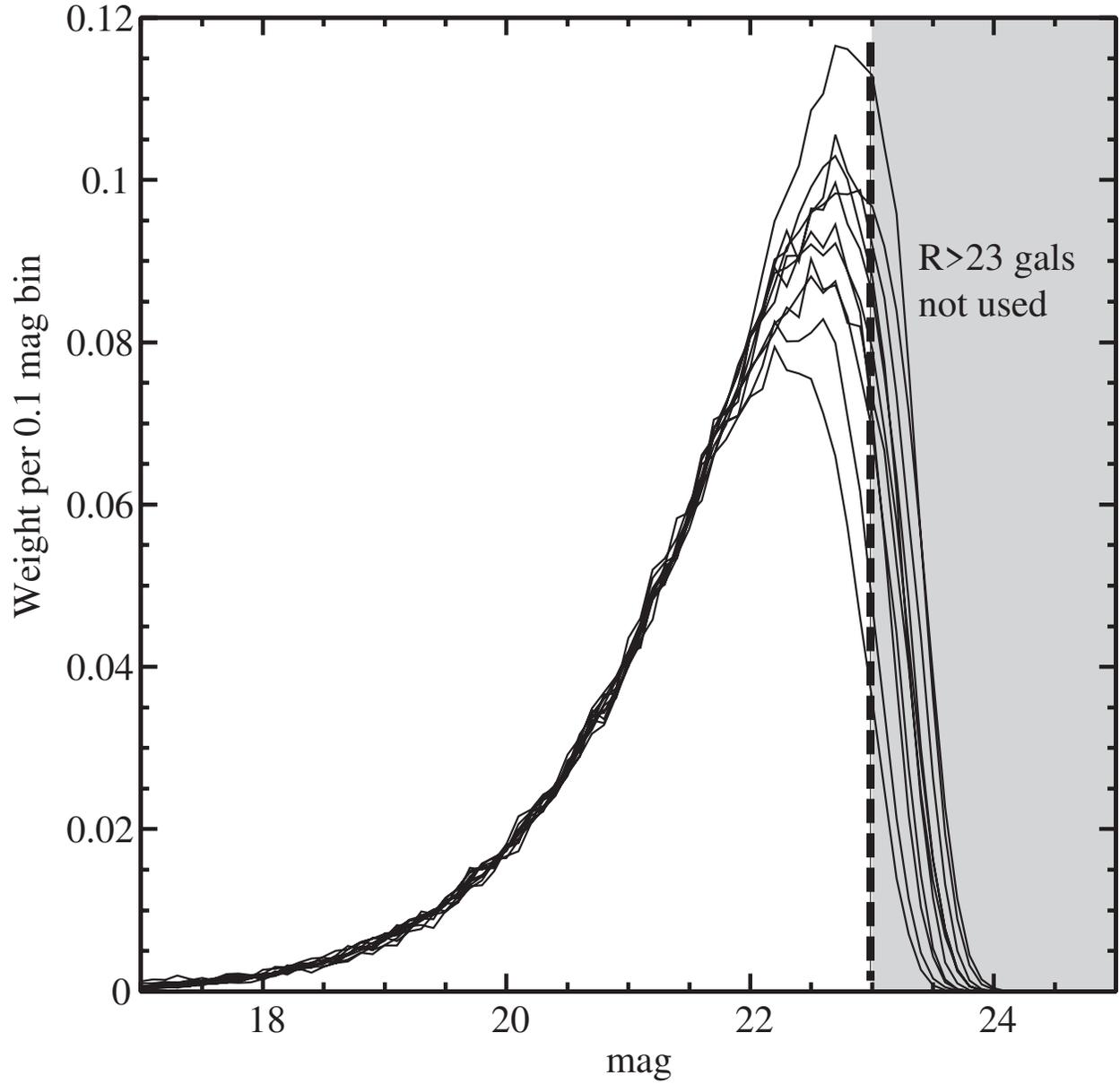}
\caption[]{ \small 
The total relative weight of our galaxies for each of our 12 fields as
a function of magnitude.  The weight function is quite similar up to 
$m=22$, at which point they begin to diverge.  We cut our catalogs
at $m=23$ to minimize variations in depth from field to field.
}
\label{weightvsmagplot}
\end{figure}

Figure~\ref{weightvsmagplot} plots the distribution of weight vs. magnitude for
galaxies which pass the selection criteria in step~\ref{stepuber}.  Each of
the 12 fields are plotted separately.  We truncate the galaxy sample
at $R<23$ to minimize field-to-field variations in effective depth,
and to keep the galaxy sample to the magnitude range for which sizable
redshift surveys are available.  We also truncate at $R>19$, since 
bright galaxies are negligibly lensed.

The Caltech Redshift Survey
\citep{Coh00}[CRS] is almost complete for ${\cal R}=23.5$ for a 437 galaxy
sample surrounding the Hubble Deep Field.  The \citet{Ste92}
${\cal R}$ photometric
system (not to be confused with our symbol for responsivity) differs
slightly in both zeropoint and bandpass from the $R$ system defined by
\citet{Lan92}.  Convolution of synthetic galaxy spectra at a
variety of redshifts (R. Somerville, private communication) suggests
that $\langle R- {\cal R}\rangle\approx 0.18$~mag for $z<0.3$ or $z>1$
galaxies, rising to $\langle R- {\cal R}\rangle\approx0.37$~mag at
$z\approx0.7$ as the 400~nm break in galaxy spectra moves between the
two filters.  We apply this correction to the CRS ${\cal R}$
magnitudes to define an $R<23$ sample that is 97\% complete.  The CRS
magnitudes are then corrected for Galactic extinction in the same way as
our program fields.

We apply \eqq{zsum} to this CRS sample, assuming that either (a) all the
galaxies with unknown redshift are at $z=1$, or (b) galaxies with
unknown redshift have the same $z$ distribution as other galaxies of
similar magnitude.  For the underlying weight distribution, we take
that of the field of median depth, field T.  We also examine the
effect of taking the shallowest (A) and deepest (N) fields.  Choosing
case (a) or (b) for incompleteness, or fields T, A, or N, changes the
expected \mapsq\ signal by at most 4\% from the
canonical case.  The implied uncertainty in $\sigma_8$ is, at 2\%,
insignificant compared to the measurement errors.

A calibration of our signal can also be done using
the Canada-France Redshift Survey \citep{Li95}[CFRS],
which is complete to $I_{AB}<22.5$.  While
$R$-band magnitudes are not measured in the CFRS, \citet{Sm01}
collect $R$-band images in the CFRS fields so that a
$R<23$ sample of 783 galaxies can be defined, after correcting for
Galactic extinction in the CFRS fields.  This $R<23$ subset of
the CFRS sample is not quite representative of our source galaxies
because the bluer galaxies at $R\approx23$ do not make the
$I_{AB}<22.5$ cut of the CFRS sample.  The CFRS $R<23$ sample is also
only 88\% complete in redshift.  The incompleteness in redshift and
the depth/color mismatch make the CFRS data less reliable for our
purposes than the CRS.  Nonetheless, treating the incompleteness by
method (a) above ($z=1$) gives a calibration within 5\% of the nominal
CRS case.  Using method (b), however, gives an expected \mapsq\ 
signal that is 15\% below that of the nominal CRS
value.  Given the known shortcomings of the CFRS sample, we will take
this as a very conservative 95\% CL bound on the possible error of the
CRS depth 
calibration.  Since $\mapsq \propto \sigma_8^{2.2}$
at our larger scales, the 95\% CL error on $\sigma_8$ is 7\%, to be
added in quadrature with the statistical errors and B-mode corrections.

\begin{figure}[t]
\epsscale{1.0}
\plotone{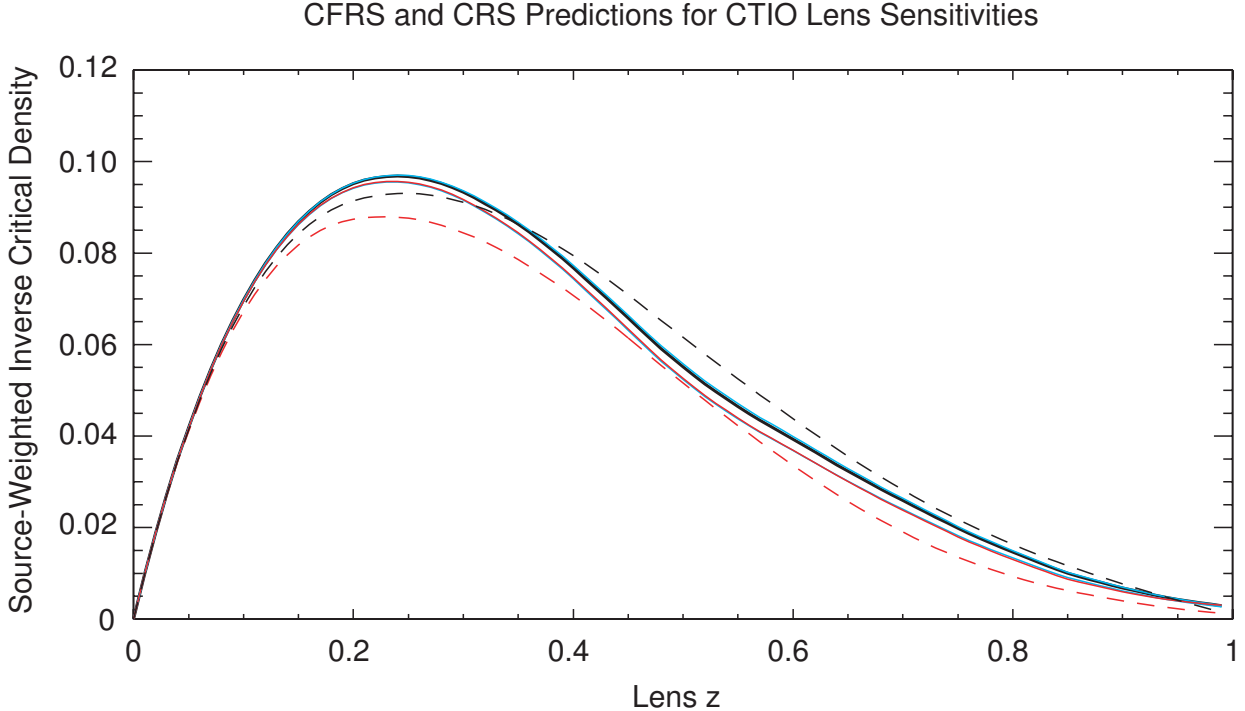}
\caption[]{ \small
Sensitivity of our survey to mass fluctuations, plotted as a
function of the lens redshift.  The underlying source distribution is
inferred from the Caltech Redshift Survey (solid) or Canada-France
Redshift Survey (dashed) as detailed in the text.  Galaxies with unknown
redshifts in these surveys are assumed to be either at $z=1$ (upper, black)
or distributed as the other galaxies (lower, red).
The differences between these curves correspond to a maximum of 7\%
uncertainty on the power spectrum normalization $\sigma_8$.
}
\label{zdistplot}
\end{figure}

There is an
additional uncertainty due to our assumptions that ${\cal R}$ and
$P(w|m)$ are independent of redshift.  Note that this assumption is
implicit in all previous cosmic-shear measurements.  A detailed test
of this assumption requires larger redshift surveys, and we will for
the time being ignore the effect as we believe it is weak.

\subsection{Prediction of Signals}
\label{models}

Using the notation of \citet{Sch98}, and approximating the source
distribution as a set of weighted delta functions at the observed
redshift-survey $z$'s,
the angular power spectrum of the
lensing convergence is expected to be
\begin{equation}
\label{pspec}
P_\kappa(k) 
= \frac{9}{4} \Omega_0^2 \int_0^{w_H}
  \frac{dw}{a^2(w)}
  P_{3D} \left( \frac{k}{f_K(w)}; w \right)
  \left[ \sum_i W_i \frac{f_K(w_i - w)}{f_K(w_i)} \right]^2. 
\end{equation}
Here $w$ is the conformal distance from $z=0$, $w_H$ is the horizon,
$w_i$ are the distances to the redshift-survey galaxies,
$f_K$ is the comoving angular diameter distance, and $P_{3D}$ is
the mass power spectrum at a given comoving scale and epoch.
Figure~\ref{zdistplot} plots the bracketed quantity in \eqq{pspec}
[with an additional factor of $f_K(w)$] as a
function of the lens redshift $z_l$, for several of the
assumptions about source-galaxy redshift distributions detailed
above---they are all seen to give very similar results.  The
sensitivity function has a broad peak at $z\approx0.25$.

There are two important issues in making model predictions for
the \mapsq\ and \gamsq\ statistics. First, for the angular scales
of interest, nonlinear clustering and its evolution must be 
taken into account.  \citet{Jai97} show that the effect
of nonlinear enhancement exceeds a factor of 4 in the variance on 
arcminute scales, and is about 50\% at 10\arcmin\ (for $\sigma_8$
close to unity). Thus essentially all existing shear-correlation 
measurements in the literature probe the nonlinear regime. 
Since our survey measurements extend to beyond 1 degree
(albeit at shallower depth), we span
an interesting dynamic range that includes the nonlinear regime
(below 5 arcminutes), the quasilinear (10-20 arcminutes) 
and the nearly linear regime on larger scales. In any case, to compare 
our full set of measurements to model predictions with some degree of accuracy
requires that we use a well calibrated model for the nonlinear 
mass power spectrum. The fitting formulae developed by 
\citet{Ha91,Pea94,Jai95,Pea96} provide empirical but fairly accurate
predictions for the nonlinear power spectrum. We will use the 
\citet{Pea96} formulae to compute the shear variances 
for different models. Recently vW02 have discussed
the accuracy achieved with these formulae and found that on arcminute
scales there is some uncertainty in the theoretical predictions
which precludes parameter estimation at much better than the 10\% 
percent level.  The large angular scale coverage of our 
measurements makes this uncertainty on small scales less of an issue. 

The second issue in making model predictions is the choice of 
cosmological parameters to vary. One approach is to choose
a physical model and vary parameters that have specific meaning
within such a model. Our approach however will be closer to an empirical
one in which we will choose the parameters that lensing is most sensitive
to and take other parameters to be unknown or fixed. Based on earlier
theoretical work \citep{Ka92,BeF97,Jai97},
we choose the primary parameter space to be
that of $\sigma_8$, such that the amplitude of the 
linear power spectrum is 
$\propto \sigma_8^2$, and $\Omega_m$, the present mean mass density 
parameter of the universe. We will parameterize the shape of the power
spectrum by the standard $\Gamma$ parameter, which has a specific
physical meaning for cold dark matter models. We will fix $\Gamma=0.21$
and the primordial spectral index $n=1$ 
in most of the analysis, but as discussed below, we will explore the
sensitivity of our constraints to the value of $\Gamma$. We have above
explored the sensitivity to the uncertainty in our redshift distribution. 
Our analysis is similar in spirit to that of vW02, but our choice of
survey depth and angular scale makes the interpretation insensitive to
unmeasured parameters of $N(z)$ and $\Gamma$, so marginalization over
these quantities is not required. 

The theoretical predictions for \gamsq\ and \mapsq\ are given by
the following equations:
\begin{align}
\label{theoryvar}
\langle \gamma^2(\theta)\rangle 
&= \frac{2}{\pi\theta^2}\int_0^\infty 
   \frac{dk}{k} P_\kappa(k) J_1(k\theta)^2,  \\
\label{theorymap}
\langle M_{\rm ap}^2(\theta)\rangle 
&= \frac{288}{\pi\theta^4}\int_0^\infty 
   \frac{dk}{k^3 } P_\kappa(k) J_4(k\theta)^2,
\end{align}
where $J_1$ and $J_4$ are the first and fourth-order Bessel functions, and
$P_\kappa$ is given by \eqq{pspec}. 
In the linear regime, for an Einstein-de Sitter universe
the 3-dimensional mass power spectrum grows
in time as $a^2$, so the dependence on the redshift coordinate $w$ is 
contained in the term in square brackets in \eqq{pspec}. 
However for other cosmologies and in the nonlinear regime the growth 
rate is different, so the dependence on $w$ is more complicated
and can be scale dependent. Thus the dependence on cosmological
parameters enters in rather complicated ways through the distance
factors as well as the power spectrum. For a reasonable class of
models that are at all consistent with other cosmological probes, 
the dependence of the second moment of lensing statistics on 
the cosmological constant is weak. The main dependences then are
on $\Omega_m$ and $\sigma_8$, and as shown in previous work, the
combination that enters is close to the one in cluster abundances
and is given roughly by 
$\langle\gamma^2\rangle\propto \sigma_8^2\Omega_m$.
There  are further
dependences on the shape of the power spectrum and the redshift
distribution which we will explore below. 

\subsection{Fit of Models to the Data}
\label{fitmodels}

As described in \S\ref{covarmatrix} above, our data are essentially
reduced to a mean vector, ${\bf x}$, of \mapsq\ and \gamsq\ values,
along with the covariance matrix of these values, \bfS.  
We calculate the corresponding vector for each of our cosmological models
and compute a $\chi^2$ value for each model:
\begin{equation}
\chi^2 = \left( {\bf x_{data}} - {\bf x_{model}} \right) \bfS^{-1}
\left( {\bf x_{data}} - {\bf x_{model}} \right)
\end{equation}

One important consideration in the above calculation is that our 
covariance matrix is largely degenerate.  Our vector has over
100 elements, since we calculate the statistics at fairly small
separations in aperture radius.  However, as discussed in 
\S\ref{covarmatrix} and \S\ref{bmode},
there are really only 3--4 independent data points among all of these.
Rather than selecting 4 values of $\theta$ arbitrarily, we 
use a singular value decomposition to calculate $\bfS^{-1}$,
and take only the largest 4 eigenvalues of the decomposition.
This automatically uses only the non-degenerate components
of the matrix, and in some sense finds the best combination
of the data to use for our 4 values.  The 4 values selected for
Table~\ref{datatable} are therefore merely representative of the 
data used for the constraints.

\begin{figure}[t]
\epsscale{1.0}
\plotone{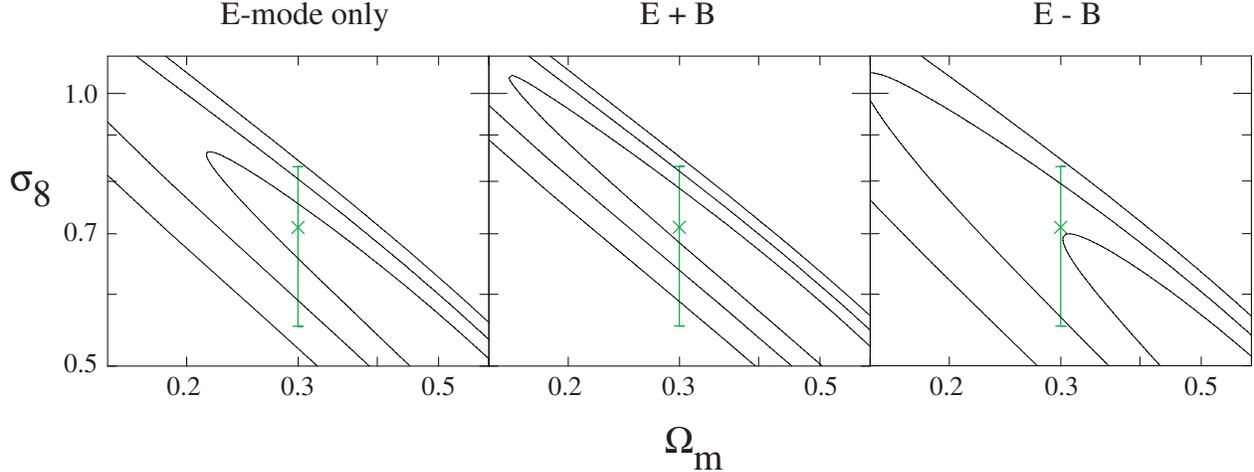}
\caption[]{ \small 
Contour plots of $\chi^2$ of our data as a function of $\Omega_m$
and $\sigma_8$, assuming $\Lambda$CDM, $\Gamma = 0.21$. 
The left plot uses the data as measured, taking only the E-mode
signal for the \map\ statistic.  For the middle plot, the B-mode
signal has been added to the E-mode.  And for the right plot, the
B-mode signal has been subtracted from the E-mode.
Contours in all cases are drawn at 1, 2, and 3 sigma, which correspond to 
$\Delta\chi^2$ = 2.30, 6.17, and 11.8 for two parameters.
Marginalizing down to one parameter yields tighter 1-dimensional 
uncertainty intervals.  We also include our final 95\% confidence 
range for $\Omega_m = 0.3$ (from \protect\eqq{answer})
on each plot for reference.
}
\label{contourplot}
\end{figure}

Figure~\ref{contourplot} shows a contour plot of $\chi^2$ 
as a function of $\Omega_m$ and $\sigma_8$. 
The nominal overall best fit is at 
\(\sigma_8 = 0.37, \Omega_m = 0.82\).
However, there is clearly
a strong degeneracy in this plot indicating that we really
only constrain the combination $\sigma_8 \Omega_m^{0.57}$, 
which is found to have a value of $0.334 \pm 0.040$ 
(95\% confidence interval).  

This error bar includes only the statistical and cosmological 
variations which went into the calculation of the covariance
matrix.  
However, we need to account for the systematic uncertainty due to 
the B-mode power.
As discussed above in \S\ref{bmode}, 
we allow for the two possibilities: that the 
E-mode should be either increased or decreased by the amount
of the B-mode.  For these two extremes (also plotted
in Figure~\ref{contourplot}), we find that 
$\sigma_8 \Omega_m^{0.57}$ ranges from 0.263 to 0.377 at the 
95\% confidence level.  That is, these are the minimum and 
maximum values at 95\% confidence for any of the three cases:
E, E+B or E-B.
These values then span the full range allowed
by our data taking into account both the systematic and the 
statistical errors.

Our calibration and redshift uncertainties of ~5\% and ~7\% respectively 
are smaller than the above errors but need to be included.
We also examined the dependence of our result on $\Gamma$, and found 
that for the range [0.15,0.50], our estimate of $\sigma_8$ varied as
$\Gamma^{-0.02}$, which even for the extremes of this range is only a 
2\% result, and is therefore negligible compared to our other uncertainties.

We add all of these uncertainties in quadrature to get a final estimate of:
\begin{equation}
\label{answer}
\sigma_8 \left(\frac{\Omega_m}{0.3}\right)^{0.57} = 0.71^{+0.12}_{-0.16} 
\quad \hbox{(95\% CL)}
\end{equation}
which includes all systematic, statistical, and calibration uncertainties.

Models for the high and low extremes of this range along with the
best fit model are the dotted curves plotted in
Figure~\ref{mapgammaplot}.

\subsection{Potential Causes of the B-Mode}
\label{potentialcauses}

It is particularly unfortunate that we cannot use the data
at $\theta<30\arcmin$, since one of the expected benefits
from our survey was the large dynamic range over which we are
able to measure shear.  We make a significant lensing detection from
$1\arcmin$ up to $\gtrsim150\arcmin$, over 2 decades of power-spectrum
range.  In the absence of contaminating power, our uncertainty in
$\sigma_8$ would be approximately half of our present error bar.
The benefit of having a larger scale range is that it can break
the degeneracy seen in Figure~\ref{contourplot} between
$\sigma_8$ and $\Omega_m$.  Larger values of $\Omega_m$ tip
the predicted curve up at small scales, whereas smaller values
tip it down.  
So with the full range of data and no B-mode contamination,
we would start to gain some constraint on $\Omega_m$.

If we repeat the above calculation using the entire range of 
1\arcmin--100\arcmin\ for the \mapsq\ statistic and the same
50\arcmin--100\arcmin\ for the \gamsq\ statistic, we find that
the statistical error bars drop from about 17\% to 
about 7\%.  However, the systematic errors dominate in this
case, so that the final 95\% CL estimate is 
$\sigma_8 (\Omega_m/0.3)^{0.47} = 0.75^{+0.23}_{-0.17}$.
Thus, while the statistical precision is nominally improved, 
the B-mode degrades the expected accuracy, and therefore we believe the
above estimate (\eqq{answer}) is more appropriate.

Given the obvious benefits to removing whatever is causing the B-mode
contamination, we have spent considerable time trying to determine
the cause and which steps have brought the B mode down to its current
level. A detailed discussion is in \citet{Jar02a}, which we summarize 
here. 

One step which we believe is a source of spurious power
is the fit of the kernel across the image (\S\ref{datareduction}, 
step~\ref{stepconvolve}).  We found the B-mode to drop somewhat
when we switched from a polynomial fit to a smoothing spline.
There are unfortunately only $\approx100$ stars per image, which
means that even without noise, a fit can only probe variations on
scales larger than $\approx 1/10$ of the chip size ($\approx 1.5\arcmin$).  
Smoothing the PSF fit to average measurement noise means that
real variations in the PSF on scales of several arcminutes will be 
missed.\footnote{
Indeed the sparseness of PSF test points, a.k.a. stars, may be the
ultimate limitation for ground-based weak lensing measurements on
aperture scales of a few arcminutes or below (A. Refregier, private
communication).  Orbiting observatories, with PSFs that are stable
over time, can accumulate a PSF map from the stars of many exposures. 
}

Another source of spurious power could be higher-order asymmetries
left in the PSF after application of our circularization kernel.
We currently use a $7\times7$ kernel
which removes the lowest 3 orders of the PSF bias.  
Stars in our fields do have significant power in higher orders.
We tried to test this 
hypothesis by increasing our kernel size from 
$5\times5$ to $7\times7$
(which we still use here), but the B-mode did not change 
significantly.  

\begin{figure}[p]
\epsscale{0.63}
\plotone{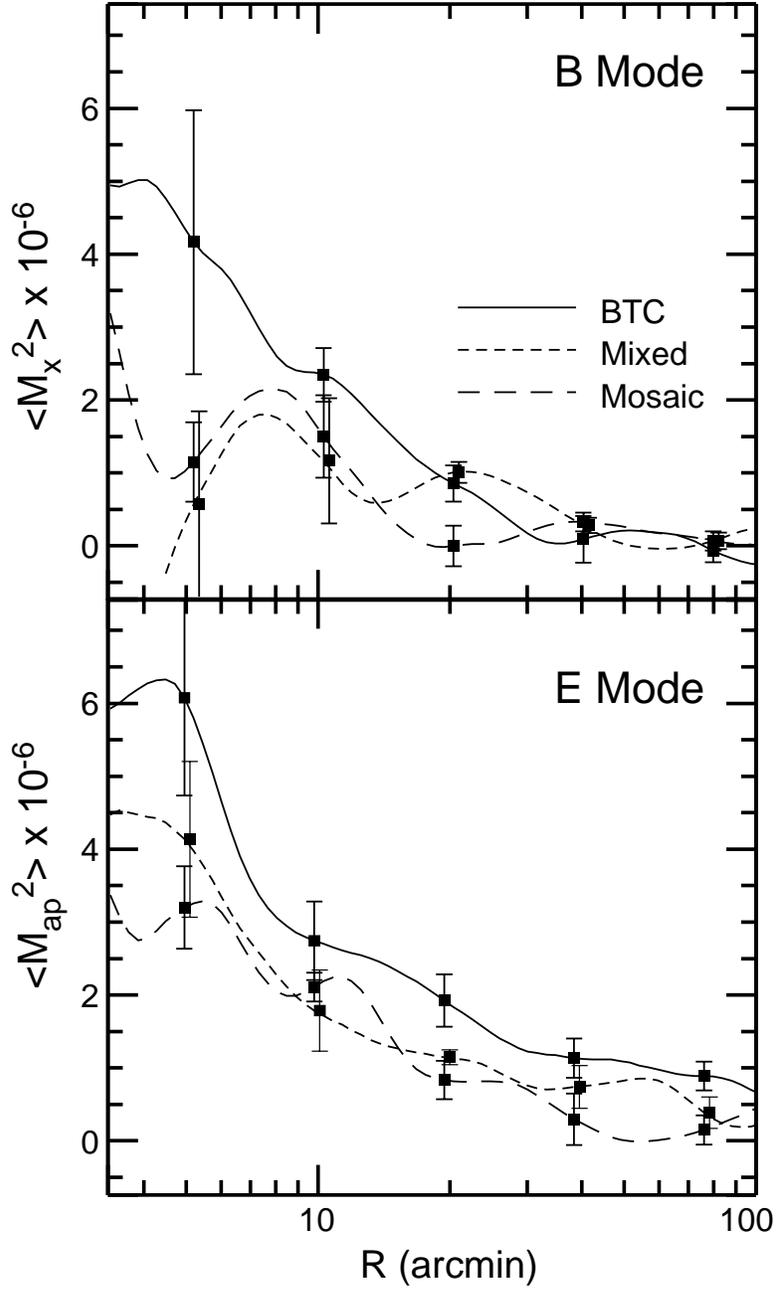}
\caption[]{ \small 
E and B mode curves for the 4 fields which were taken with the
BTC camera, the 4 taken with the Mosaic camera, and the 4 mixed fields.
The three types of fields are seen to be generally consistent with each 
other, although there is some indication that the BTC fields may have 
slightly more power.  It is also
important to note that none of the three types of fields have significant
B-mode power at $\theta>30\arcmin$.
}
\label{mapeb}
\end{figure}

Non-linearities in the CCD response could produce spurious shear
power,  because we use
relatively bright stars to measure the size and shape of the
PSF.  If this were the case, however, the effect would likely
be very different for the two cameras used in our observations.
The measured E and B powers for BTC, Mosaic, and mixed
fields are all consistent with each other, as illustrated in
Figure~\ref{mapeb}. 

While we do not know which of the above effects is causing our
B-mode, shortcomings in the PSF-circularization kernel
seem the most likely cause.  We may
be able to do better by implementing the analytic 
deconvolution technique described in BJ02.

To be complete, there is another possible cause of the
B-mode contamination---intrinsic correlations of the galaxy
ellipticities.  \citet{Cr01} calculate the correlation function
that may exist due to spin correlations between galaxies.
The predicted power, scaled
to our median redshift of 0.5, is not much lower than the B-mode
power seen in our data.  They indicate in their conclusions that 
there are a number of reasons to suspect that the predictions are
an overestimate of the true power from intrinsic correlations;
however, it is possible that a large portion of our B-mode power
is due to this effect.  Note that if this is the case, then 
\citet{Cr02} calculate that the power should be approximately
evenly split between the E-mode and B-mode, in which case, the 
E-B fit above is the appropriate one for weak lensing constraints.

\subsection{Consumer's Guide to Current Weak Lensing Results}
\label{comparetoexisting}

Table~\ref{lensingtable} lists some of the most recent weak-lensing
results, quoting the value of $\sigma_8$ 
at $\Omega_m = 0.3$ and $\Gamma = 0.21$ 
in order to make the comparisons more
obvious.  Note, however, that all authors actually
constrain a parameter $\sigma_8 \Omega_m^\alpha \Gamma^\beta$,
where $\alpha$ is usually $\approx$ 0.4--0.6, and $\beta$ 
ranges from -0.02 for our survey (\cf \S\ref{fitmodels}) to
0.15 for \citet{Re02}.
Our result is clearly somewhat lower than the other recent
$\sigma_8$ values, but the relatively large error bars
for most of these
are such that we are nominally consistent with all
of them (although we are only barely consistent with
vW02 at 95\% confidence).

\begin{deluxetable}{lcccccccc}
\tabletypesize{\footnotesize}
\tablecolumns{9}
\rotate
\tablewidth{0pt}
\tablecaption{Recent Weak Lensing Measurements of $\sigma_8$}
\tablehead{
\colhead{} &
\colhead{$\sigma_8$\tablenotemark{1}} &
\colhead{Mag Limit} &
\colhead{${\bar z}$} &
\colhead{Max scale} &
\colhead{Survey Area} &
\colhead{B Mode?} &
\colhead{$N(z)$} &
\colhead{$\Gamma$}
\\
\colhead{} &
\colhead{(95\% CL)} &
\colhead{} &
\colhead{(source)} &
\colhead{(arcmin)} &
\colhead{(sq. deg.)} 
}
\startdata
This Work & 
    $\mbox{0.71}^{+\mbox{\scriptsize 0.12}}_{-\mbox{\scriptsize 0.16}}$ & 
    $R<23$ & 0.66 & 100 & 
    75 & yes & 
    spectroscopic & irrelevant \vspace{3pt} \\ 
\citet{Ha02}\tablenotemark{8} &
    $\mbox{0.69}^{+\mbox{\scriptsize 0.41}}_{-\mbox{\scriptsize 0.22}}$ &
    $I<24.5$ & $\sim 1$ & 40 & 
    2.1 & yes & 
    marginalized\tablenotemark{4} & 
    marginalized\tablenotemark{7} \vspace{3pt} \\ 
\citet{Br02}\tablenotemark{8} &
    0.74\mbox{\scriptsize$\pm$0.18} & 
    $R<24$ & 0.85 & 15 & 
    1.25 & yes & 
    photometric &
    $\Gamma = 0.72 \Omega_m$ \vspace{3pt} \\ 
\citet{Ho02}\tablenotemark{2} & 
    $\mbox{0.86}^{+\mbox{\scriptsize 0.08}}_{-\mbox{\scriptsize 0.10}}$ &
    $R<24$ & 0.58 & 50 & 
    53 & yes & 
    marginalized\tablenotemark{3} & 
    marginalized\tablenotemark{6} \vspace{3pt} \\ 
\citet{vW02} & 
    0.96\mbox{\scriptsize$\pm$0.23} & 
    $I<24.5$ & 0.84 & 30 & 
    8.5 & yes & 
    marginalized\tablenotemark{4} & 
    marginalized\tablenotemark{7} \vspace{3pt}\\ 
\citet{Bac02} & 
    0.97\mbox{\scriptsize$\pm$0.26} & 
    $R<25.8$ & 0.9 & 12 & 
    1.6 & unknown & 
    photometric\tablenotemark{5} & 
    $\Gamma = 0.21$ \vspace{3pt}\\ 
\citet{Re02} & 
    0.94\mbox{\scriptsize$\pm$0.28} & 
    $I<24.5$ & 0.95 & 1.4 & 
    0.04 & unknown & 
    photometric\tablenotemark{5} & 
    $\Gamma = 0.21$
\enddata 
\tablenotetext{1}{ Assuming $\Lambda$CDM model, 
normalized to $\Omega_m = 0.3$, $\Gamma = 0.21$. }
\tablenotetext{2}{ See \S\ref{comparetoexisting} 
for his preliminary results using the \protect\citet{Sm02} formulation
of the non-linear power spectrum. }
\tablenotetext{3}{ A functional form for N(z) is taken from spectroscopic
surveys.  The fit is marginalized over a free parameter, $z_s$ of this 
function, with a Gaussian prior for $z_s$ based on photometric
redshifts. }
\tablenotetext{4}{ Same as (3), but with a flat prior for $z_s$. }
\tablenotetext{5}{ The median redshift is estimated from a photometric
extrapolation of spectroscopic surveys. }
\tablenotetext{6}{ Assuming a Gaussian prior based on the 2dF galaxy
redshift survey. (With a flat prior, the estimate of $\sigma_8$ 
increases to 
$\mbox{0.91}^{+\mbox{\scriptsize 0.10}}_{-\mbox{\scriptsize 0.24}}$.)}
\tablenotetext{7}{ Assuming a flat prior [0.1,0.4]. }
\tablenotetext{8}{ The results of \citet{Br02} and \citet{Ha02} 
were published as preprints while this paper was being reviewed.
We have added a summary of their results here, although they are not
discussed in the text of \S\ref{comparetoexisting}. }
\label{lensingtable}
\end{deluxetable}

Direct comparison of various results is complicated, however,  because 
the uncertainties included in the error bar vary, even though
all the error bars are 95\% confidence.\footnote{
When quoted error bars were 1-sigma, we
doubled the uncertainty to get the 95\% confidence limits.}
We note here some of the important distinctions between current
results which must be considered when making detailed comparisons.
Table~\ref{lensingtable} serves as a ``scorecard'' for recent
cosmic-shear results.

\begin{itemize}

\item{\bf Statistical and Cosmic Variance:}
Uncertainties in shear power estimates of all the papers in
Table~\ref{lensingtable} include the contributions of random (shape)
noise and cosmic variance.  For vW02 and RRG02 these are estimated
analytically.  The other authors use measured field-to-field
covariance matrices, which will automatically include shape noise and
cosmic variance.

\begin{figure}[t]
\epsscale{1.0}
\plotone{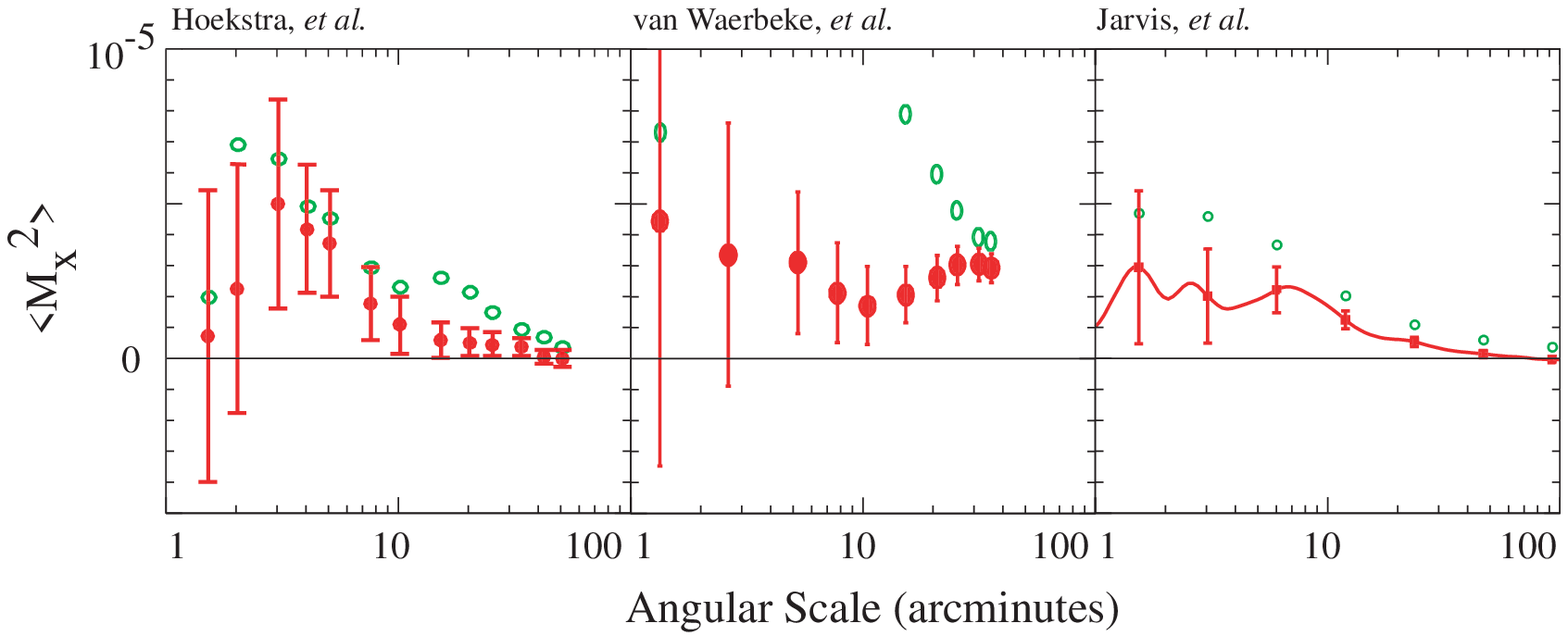}
\caption{ \small
Measurements of the B-mode contamination from each of the three
surveys which have tested for it so far.  The left panel is
taken from Figure~1 of HYG02 (for their entire magnitude
range $20<R<24$; the data restricted to $22<R<24$ have slightly
lower B-mode power).  The middle panel is
taken from Figure~1 of vW02.  The final panel is
taken from Figure~\protect\ref{mapgammaplot}, and converted into linear
rather than logarithmic scale for the y-axis
to match the other authors.  For all three plots, the green open circles
are the E-mode measured for that study.
}
\label{hovwplots}
\end{figure}

\item{\bf Estimates of Spurious Power:} 
``Cosmic shear'' measurements are difficult, since spurious sources of
shear variance are easily mistaken for lensing signals.  The E/B
decomposition is one test for the presence of spurious power.
HYG02 and vW02 are the only other authors to 
look for (and find) B-mode in their data; the RRG02 and BMRE02 surveys
have field sizes too small for a useful E/B decomposition.
Figure~\ref{hovwplots} compares the B-mode power
detected in the three relevant efforts, which is significant in all
cases, so must be included in the error analyses.

Both vW02 and HYG02 add the B-mode signal 
to the {\em error bar} of the E-mode
signal (in quadrature).  This implicitly assumes that the effect of the 
systematic error is independent (or weakly correlated) at different
angular scales.  It seems more likely to us that
the direction of the error is
highly correlated at all scales, in which case the HYG02/vW02
procedure could underestimate the systematic uncertainty---hence our
approach of subtracting (or 
adding) the B mode from the {\em signal} of the E-mode at all angular
scales in order to bound systematic errors.
The analysis technique of vW02 and HYG02
could underestimate the systematic uncertainty.
See \citet{Jar02a} for further discussion of the B-mode
signals present in our, HYG02, and vW02 data.  

\item{\bf Source Redshift Distribution:}
Inaccuracies in the assumed $N(z)$ for source galaxies will cause
scale errors in the derived $\sigma_8$.  In our shallow survey, $N(z)$
and its uncertainties can be taken directly from nearly-complete
spectroscopic galaxy surveys.  The other papers in
Table~\ref{lensingtable} assume a parametric 
form for $N(z)$, with parameters fit to the measured median photometric
redshifts in the Hubble Deep Field (and other data).  HYG02 and vW02
then marginalize the resultant $\sigma_8$ over an estimated prior
distribution for the $N(z)$ parameter.  RRG02 and BMRE02 fix the
median redshift based on their limiting magnitudes, but propagate the
resultant uncertainty into their uncertainty in $\sigma_8$.

\item{\bf Power Spectrum Shape:} 
Cosmic shear studies currently provide 
useful constraint only on the overall normalization of the mass power
spectrum, not its shape.  If the shear measurement is at scales far
from the $\approx 8$~Mpc where the spectrum normalization $\sigma_8$
is defined, then $\sigma_8$ will depend upon the assumed shape of the
power spectrum.  The (linear) CDM power spectrum is specified by the
primordial index $n$ and the parameter $\Gamma$.  All papers to date
assume $n=-1$; vW02 and HYG02 marginalize over a prior for $\Gamma$.
For our results, the measurement is on larger physical scales, near
the window for $\sigma_8$ itself; this means that our $\sigma_8$
result is very weakly dependent upon $n$ and $\Gamma$.

Note that in the current results listed in Table~\ref{lensingtable},
the trend is that surveys using smaller scale tend to have larger
measured values for $\sigma_8$.  
If taken at face value, this would imply that $\Gamma$ is larger than the
usually assumed value of 0.21.  However, this is at best suggestive,
and shouldn't be taken too seriously yet.

\item {\bf Non-Linear Mass Evolution:} 
Constraint of cosmological
parameters with weak lensing data requires a model for non-linear
evolution of the power spectrum.  Various authors have put forth
heuristic methods for estimating non-linear evolution, but these have
not been verified (at the $\approx10\%$ level) for the range of
cosmologies considered
here, and the baryonic contribution to the small-scale power spectrum
is also significant when accuracies of a few percent are desired.
\citet{Pea96} claim that their prescription
for non-linear predictions
has an accuracy of 15\%, which is roughly
half the discrepency between the studies using small angular scales
and our results.  

Further, Hoekstra has redone the analysis for the RCS survey using
the new \citet{Sm02} formulation for the non-linear evolution. 
He obtained a value of 0.80 for $\sigma_8$ which is about 8\% lower than 
the published value which uses the formulation of \citet{Pea96}
(preliminary results; Hoekstra, private communication), indicating that
the other results relying on the non-linear regime may also be biased
high.
This potential source of systematic error is reduced for
larger-scale surveys such as this work.

\end{itemize}

There seems to be a trend in Table~\ref{lensingtable} wherein the 
lower $\sigma_8$ values are obtained by the shallower, larger-scale surveys
(ours and HYG02).  This may indicate that mis-estimation of $N(z)$ and
non-linear evolution are causing biases on $\sigma_8$ for the deeper,
smaller-scale surveys.  We also note that it is easier for uncorrected
PSF variations to falsely inflate $\sigma_8$ than to decrease it.
We would not, however, choose to
over-interpret these apparent discrepancies until such time as larger
datasets that are free of B-mode become available.

A close comparison between our result and HYG02 is illustrative given
the similarities of the datasets.  The weight of the HYG02 data
peak near $R=23.5$ mag, only $\approx0.5$~mag deeper than our data, so
we expect the $\mapsq$ signals to agree within $\approx20\%$, but
their higher $\sigma_8$ result implies that they have measured a
$\mapsq$ more than 1.5 times ours.  At
large scales, $\theta>30\arcmin$, the HYG02 signal is in fact lower than, but
consistent with ours, whereas at the intermediate scales that we
ignore, the HYG02 signal
is higher.  Hence the $\sigma_8$ differential between the results is
partly due to different scales that we have fit.   As noted above, a
change in the non-linear growth model reduces the difference between
the HYG02 $\sigma_8$ and ours.\footnote{
We also note that
the $N(z)$ assumed by HYG02 for their sample has a {\em lower} mean
$z$ than we have assumed, despite their {\em deeper} sample.  This
discrepancy in $N(z)$ models accounts for roughly half of the
difference in resultant $\sigma_8$. }

\subsection{Comparison to X-Ray Cluster Measurements}

\begin{deluxetable}{ll}
\tablewidth{0pt}
\tablecaption{Recent Cluster Measurements of $\sigma_8$}
\tablehead{
\colhead{ } &
\colhead{$\sigma_8$\tablenotemark{1}} \\
\colhead{ } &
\colhead{(95\% CL)}
}
\startdata
\citet{Pi02} & 0.77\mbox{\footnotesize$\pm$0.10} \vspace{3pt} \\
\citet{Bah02} & 0.72\mbox{\footnotesize$\pm$0.14} \vspace{3pt} \\
\citet{Vi02} & 0.61\mbox{\footnotesize$\pm$0.09} \vspace{3pt} \\
\citet{Se02} & 0.75\mbox{\footnotesize$\pm$0.12} \vspace{3pt} \\
\citet{RB02} & 0.68\mbox{\footnotesize$\pm$0.11} \vspace{3pt} \\
\citet{Bo01} & 0.72\mbox{\footnotesize$\pm$0.12} \vspace{3pt} \\
\citet{PW01} & 0.87\mbox{\footnotesize$\pm$0.13} \vspace{3pt} \\
\citet{Si01} & 1.01\mbox{\footnotesize$\pm$0.18} \vspace{3pt} \\
\citet{Pi01} & 1.01\mbox{\footnotesize$\pm$0.14}
\enddata
\tablenotetext{1}{Assuming $\Lambda$CDM model, normalized to $\Omega_m=0.3$, $\Gamma=0.21$.}
\label{clusterstable}
\end{deluxetable}

It is also interesting to compare these weak lensing results
with the recent results from X-ray cluster measurements,
summarized in Table~\ref{clusterstable}.
The precision of the two types of measurement are 
comparable, but the cluster method suffers from 
very different systematics than do weak lensing 
measurements.
In particular, there is significant uncertainty
in the Mass-Temperature relation which relates
to the $\beta$ parameter of the assumed density profiles.
This systematic error is seen to dominate the
uncertainty in the $\sigma_8$ measurements, since
the scatter of the measurements is significantly
larger than the quoted 95\% confidence intervals.
(This systematic effect is discussed in more detail
by \citet{Pi02}.)

\citet{Hut02}
characterize the Mass-Temperature relation as:
\begin{equation}
\left(\frac{M(T,z)}{10^{15}h^{-1}M_\sun}\right) 
= \left(\frac{T}{T_*}\right)^{3/2} 
  \left(\Delta_c E^2\right)^{-1/2}
  \left(1-2\frac{\Omega_\Lambda(z)}{\Delta_c}\right)^{-3/2}
\end{equation}
where $\Delta_c$ is the mean overdensity inside the 
virial radius in units of the critical density, 
\(E^2 = \Omega_m(1+z)^3 + \Omega_\Lambda + \Omega_k(1+z)^2\).
In their investigation of how cluster mass measurements
constrain cosmology, they conclude that the parameter,
$T_*$ is directly related to the combination 
$\sigma_8\Omega_m^{0.6}$.  Since this is essentially 
the same as what we measure, our measurement implies
(according to their Figure~1) a value of 
$T_* \approx 1.5$~keV, which is at the high
end of their ``favored range''.  

\section{Conclusions}
\label{resultsconclusion}

We have presented the results of a 75 square degree survey of
galaxy shapes involving 12 well-separated fields and totaling
approximately 2 million galaxies.  We have applied the 
analysis techniques of BJ02
to our data, and show many tests for residual systematic 
effects.  Most of these tests show no residual effect,  
and for the one exception (\cf Figure~\ref{biasplot}), 
the bias is small compared to our lensing signal and we
correct our shapes for this bias as well.

We calculate two lensing statistics for our data, the shear
variance and the aperture mass, the results of which are plotted
in Figure~\ref{mapgammaplot}.  The presence of a so-called
B-mode in the aperture mass statistic indicates that we probably
have some residual systematic effect which is causing spurious correlations
in our galaxy ellipticities.  While we do not believe that 
it is due to intrinsic correlations between the galaxies, we
cannot rule out this possibility.

In any case, the presence of B-mode in the smaller-scale portion
of our data precludes us from using
this part of the range for constraining cosmology.  Using
the data for angular scales $\theta > 30\arcmin$, where the 
B-mode is small, we are able
to jointly constrain the parameters $\sigma_8$ and $\Omega_m$.
They are found to be largely degenerate to the extent that we 
effectively only constrain the combination:
\begin{equation}
\sigma_8 \left(\frac{\Omega_m}{0.3}\right)^{0.57} = 0.71^{+0.12}_{-0.16} 
\quad\hbox{(95\% CL)}
\end{equation}
where the error bars are 95\% confidence and include the statistical,
calibration and systematic uncertainties.  There is no dependence upon
the Hubble Parameter $H_0$, and dependence upon the power spectrum
parameters $n$ and $\Gamma$ are insignificantly small over the range
of reasonable values.

Our value for $\sigma_8$ is lower than all other cosmic shear results
to date, but formally consistent with all but one at the 2$\sigma$ level.
While it is possible that the discrepency may be due to 
systematic errors from the B-mode contamination discussed above,
it may also be related to the treatment of source redshifts and/or
non-linear mass evolution.
The other weak lensing result with which we most closely agree is
HYG02.  This study is the most similar to ours in both 
source redshifts and angular scale.  

The fairly large scales to which we have confined our analysis are
less sensitive to any systematic errors which may exist in the 
non-linear predictions.  The relatively bright source galaxies in our survey 
also allow us to minimize the potential 
systematic error due to the redshift distribution.
We are able to use existing spectroscopic redshift surveys to calibrate our 
redshift distribution, as described in \S\ref{zdist}.  

Also, our results are consistent with the latest several results using
cluster abundances, but are inconsistent with the ``older'' paradigm. 
There are clearly some systematic effects in this field which must be
worked out, but once these are worked out, 
our value of $\sigma_8$ could have significant 
consequences for the physics of clusters.
For example, as we pointed out above,
our results may imply a relatively large value for $T_*$ (as defined
by \citet{Hut02}).

Finally, as also pointed out by vW02, our likelihood contours
are roughly orthogonal to those of \citet{Lah02} from CMB+2dF constraints.
(See also \citet{Me02} for a similar investigation.)
The intersection of our contours with theirs (\cf their Figure~2) 
falls roughly at 
$\Omega_m = 0.3$, $\sigma_8 = 0.7$.  This result, along with the 
CMB result that $\Omega_m + \Lambda = 1$, thus supports the
currently popular model with $\Omega_m = 0.3, \Lambda = 0.7$.  
In particular, our results (and in fact almost all of the weak lensing 
results to date) are inconsistent with an $\Omega_m = 1$ flat model.

In a relatively short time, weak gravitational lensing measurements
have advanced to ``precision cosmology'' status, constraining at least
one parameter combination to the $\approx$ 10\% level.  Unfortunately the
{\em accuracy} of the method is currently limited by residual PSF
contamination, 
as indicated by B-mode power.  Improvements in image quality and
analysis techniques should lead to further rapid improvements in the
power of weak lensing cosmological constraints.

\acknowledgements
This work was supported by grant AST-9624592 from the National Science
Foundation.  We thank the CTIO TAC and staff for providing many of the
resources and excellent support necessary for this challenging project.
We also thank the anonymous referee for several useful comments which
have improved the paper.

\end{document}